\documentclass[10pt,journal,compsoc]{IEEEtran}

\usepackage{url}
\usepackage{booktabs}
\usepackage{calc}
\usepackage{graphicx}
\usepackage{soul,color}
\usepackage{multirow}
\usepackage{algpseudocode,algorithm,algorithmicx}
\usepackage{comment}
\usepackage{paralist}
\usepackage{color}
\usepackage{paralist}
\usepackage{xcolor}
\usepackage[nocompress,noadjust]{cite}
\usepackage{amsmath}
\usepackage{amssymb}
\usepackage{amsthm}
\usepackage{bm}
\usepackage{xspace}
\usepackage{enumitem}
\usepackage{pifont}
\usepackage[utf8]{inputenc}
\usepackage{standalone}
\ifCLASSOPTIONcompsoc
	\usepackage[caption=false, font=footnotesize, labelfont=sf, textfont=sf]{subfig}
\else
	\usepackage[caption=false, font=footnotesize]{subfig}
\fi

\newcommand{\descr}[1]{\medskip\noindent\textbf{#1:}}
\renewcommand{\paragraph}{\descr}

\graphicspath{{figures/}}

\newcommand*{\unit}[1]{\ensuremath{\mathrm{\,#1}}}
\newcommand{\dmin}{\ensuremath{d\textsubscript{min}}\xspace}
\newcommand{\iw}{\ensuremath{I_w}\xspace}

\newcommand{\bmin}{\ensuremath{\beta\textsubscript{min}}\xspace}
\newcommand{\bmax}{\ensuremath{\beta\textsubscript{max}}\xspace}

\newcommand{\stp}{\ensuremath{\Pi}\xspace}
\newcommand{\home}{{\sc{Home}}\xspace}

\newcommand{\omni}{\textsc{Omni}\xspace}

\newcommand{\scan}{\textsc{Scan}\xspace}
\newcommand{\dscan}{\textsc{Double-Scan}\xspace}
\newcommand{\hilbert}{\textsc{Hilbert}\xspace}

\newcommand{\lmat}{\textsc{Lmat}\xspace}
\newcommand{\drf}{\textsc{Drf}\xspace}
\newcommand{\drbc}{\textsc{Drb-C}\xspace}
\newcommand{\xiao}{\textsc{IoC}\xspace}
\newcommand{\lee}{\textsc{IoA}\xspace}

\newcommand{\dw}{DecaWave\xspace}
\newcommand{\ma}{MA\xspace}
\newcommand{\mas}{MAs\xspace}
\newcommand{\gd}{GD\xspace}
\newcommand{\gds}{GDs\xspace}

\newtheorem{theorem}{Theorem}
\newtheorem{lemma}{Lemma}

\newtheorem{fact}{Fact}

\newcommand{\revision}[1]{{\color{blue}{#1}}}

\renewcommand{\revision}[1]{{\color{black}{#1}}}

\begin{document}
\title{Measurement Errors in Range-Based Localization Algorithms for UAVs:\\ Analysis and Experimentation}
\author{
	Francesco Betti Sorbelli,~\IEEEmembership{Member,~IEEE;}
	Cristina M.~Pinotti,~\IEEEmembership{Member,~IEEE;}\hspace{2in}
	Simone Silvestri,~\IEEEmembership{Senior Member,~IEEE;} and 
	Sajal K. Das,~\IEEEmembership{Fellow,~IEEE}
	\IEEEcompsocitemizethanks{
		
		\IEEEcompsocthanksitem Francesco Betti Sorbelli is with the 
		Department of Computer Science, Missouri University of Science and Technology, Rolla, MO 65409, USA. 
		Part of this work was done while F. Betti Sorbelli was working for the {\em NALP-SAPR} Project in Perugia.
		\IEEEcompsocthanksitem Cristina M.~Pinotti is with the 
		Department of Computer Science and Mathematics, University of Perugia, Italy.
		\IEEEcompsocthanksitem Simone Silvestri is with the 
		Department of Computer Science, University of Kentucky, Lexington, KY 40506, USA.
		\IEEEcompsocthanksitem Sajal K. Das is with the 
		Department of Computer Science, Missouri University of Science and Technology, Rolla, MO  65409, USA. Part of this work was done while S. K. Das was Satish Dhawan Visiting Chair Professor at the Indian Institute of Science, Bangalore - 560012, India.
	}
}

\IEEEoverridecommandlockouts

\bstctlcite{IEEEexample:BSTcontrol}

\IEEEtitleabstractindextext{%
\begin{abstract}
Localizing ground devices (\gds) is an important requirement for a wide variety of applications, such as infrastructure monitoring, precision agriculture, search and rescue operations, to name a few. To this end, unmanned aerial vehicles (UAVs) or drones offer a promising technology due to their flexibility. 
However, the distance measurements performed using a drone, an integral part of a localization procedure, incur several errors that affect the localization \revision{accuracy}. 
In this paper, we 
provide analytical expressions for the impact of different kinds of measurement errors on the ground distance between the UAV and \gds. 
We review three range-based and three range-free localization algorithms, identify their source of errors,
and analytically derive the error bounds resulting from aggregating multiple inaccurate measurements. 
We then extend the range-free algorithms for improved \revision{accuracy}. We validate our theoretical analysis and compare the observed localization error of the algorithms after collecting data from  a testbed using ten \gds and one drone, equipped with ultra wide band (UWB) antennas and operating in an open field. 
Results show that our analysis closely matches with experimental localization errors. 
Moreover, compared to their original counterparts, the extended range-free algorithms significantly improve the \revision{accuracy}.
\end{abstract}
	
}

\IEEEpeerreviewmaketitle
\maketitle

\vspace{-0.1in}
\section{Introduction}
In recent years, unmanned aerial vehicles (UAVs) or drones have received increasing attention from both the research and industry community. 
Of particular interest are  applications enabled by combining UAVs with the  Internet of 
Things (IoT)~\cite{motlagh2016low,choi2015building}, such as environmental monitoring~\cite{gao2018high}, structural health monitoring~\cite{kang2018autonomous}, precision agriculture~\cite{tsouros2019data}, search and rescue operations~\cite{silvagni2017multipurpose}, and so on.

An important requirement of such applications is the ability to accurately localize 
the position of ground devices (\gds), making the collected data more meaningful.
Since it is costly to
equip each \gd in the network with a GPS module, a fixed
set of {\em anchor} devices, whose positions are known a-priori, is generally used~\cite{priyantha2003anchor}.
Moreover, given the anchors use wireless
transmissions to localize other \gds and their range is often
limited, the number of required anchors could dramatically increase with the size of the network, 
thus increasing the cost of the localization procedure. 
This problem can be solved by replacing fixed anchor devices with a single {\em mobile anchor} (\ma) equipped with a GPS unit and periodically broadcasting its position to help nearby \gds localize.

Although there exists some work in the literature on localization based on ground \mas, such as rovers~\cite{han2016survey},
relatively less has been proposed using flying \mas, like UAVs or drones~\cite{malhotra2019comprehensive}, which is the focus of this paper. 
Compared to the ground \mas, the flying \mas are able to reach remote locations, move at a faster speed, and cover a wider area than terrestrial rovers~\cite{bekmezci2013flying}. 
Due to these advantages, in this paper we concentrate on localization algorithms involving flying \mas. 
When using UAVs as \mas, the distance can be estimated wirelessly
\revision{by measuring the time of flight (ToF) between the \ma and \gds.
For distance measurements, in this paper we adopt the ultra wide band (UWB) technology~\cite{mueller2015fusing}.}
%

Localization algorithms can be broadly categorized as {\em range-free} and {\em range-based} approaches~\cite{han2016survey}.
In the range-free algorithms, the position is estimated without any type of measurements, 
but by only discovering if the \gd and  \ma are in range. Among these, the {\em radius-based} approaches assume the knowledge of transmission radius~\cite{ssu2005localization}, while the {\em radius-free} ones do not~\cite{xiao2008distributed}. Such algorithms are often based on the assumption that the antenna radiation pattern is isotropic, which is unrealistic in general. In fact, our recent works~\cite{bettisorbelli2019ground, bettisorbelli2020rangefree} have shown that the localization \revision{accuracy}  depends on the quality (pattern and radius) of the antenna, and how much they differ from the assumed isotropic pattern. 
On the other hand, in the range-based localization algorithms, the position of the \gd is estimated
by taking several measurements between it and the \ma. These algorithms are known to be more \revision{accurate} than range-free algorithms but at the cost of additional specialized hardware. For example,
the estimation of distance exploits techniques like the received signal strength indicator (RSSI), the time of arrival (ToA),
or the time difference of arrival (TDoA)~\cite{laaraiedh2011comparison}.

Now, in any range-based localization procedure, {\em measurement errors} are unavoidable, 
and can seriously impact the localization \revision{accuracy}. 
This is particularly relevant when the \ma is a drone because the measurement errors can occur while calculating the distance between the \ma and the \gd. 
The magnitude of such errors depends on the adopted technology and on the quality of the \revision{air-to-ground (A2G)} link between the \ma and the \gd. 
For example, the distance measurement error in a fully line of sight (LoS) link 
using the Wi-Fi technology is about $7$--$10 \unit{m}$; using Bluetooth it can be up to $15 \unit{m}$ while it is only $10 \unit{cm}$ using the UWB technology~\cite{www-Deca-dwm1000}. 
Additional errors may be caused by  
non-optimal weather conditions or the drone's GPS \revision{{\em accuracy}}~\cite{www-gps}. For instance, the 3DR Solo drone used in our experiments has a GPS \revision{accuracy} of $1 \unit{m}$~\cite{www-3DR-GPS}.
In general, these errors propagate when combined and projected to the ground to localize the \gds. The error propagation depends on  the specific localization technique, such as trilateration~\cite{thomas2005revisiting}, intersection of points, centroid~\cite{blumenthal2007weighted, bulusu2000gps}, and so on.

\paragraph{Our Contributions}
In this paper, we first provide bounds on various errors (e.g., instrumental error, rolling error, altitude error) impacting the estimated ground distance between the \ma and \gd. 
Then we focus on the commonly used {\em trilateration} based localization, and derive bounds on the propagation of ground distance errors on the estimated position of the \gd. Finally, we perform extensive {\em in-field} experiments to quantify the localization \revision{accuracy} of several existing state-of-the-art localization algorithms. Specifically, we consider the \drbc~\cite{bettisorbelli2019ground} range-based algorithm,
the \drf~\cite{bettisorbelli2019rangefree}, 
\xiao~\cite{xiao2008distributed}, and 
\lee~\cite{lee2009localization} range-free algorithms
extended to distance measurements, 
and two trilateration based algorithms like 
\scan~\cite{bettisorbelli2018range} and 
\omni~\cite{koutsonikolas2007path}. Our testbed uses two UWB \dw kits, namely EVK1000 kit~\cite{www-Deca-dwm1000} and MDEK1001 kit~\cite{www-Deca-dwm1001}, and a 3DR Solo drone as the \ma. 

To the best of our knowledge, ours is the {\em first work} that provides an extensive in-field evaluation on the localization \revision{accuracy} of the most relevant algorithms in the literature in real experimental settings using drones. 
Our novel contributions are summarized as follows.
\revision{
\begin{itemize}
	\item We derive bounds on various measurement errors (instrumental, rolling, and altitude) to estimate the impact on the estimated ground distance between the UAV (\ma) and \gd.
	
	\item We validate our theoretical analysis on the ground error with a simple set of static experiments using two UWB antennas. We observe the impact of measurement errors on the trilateration technique. 
	
	\item Through experiments, we comprehensively compare three range-based and three range-free state-of-the-art localization algorithms using a UAV as \ma, extending the range-free algorithms with distance measurements to significantly improve their localization accuracy.
	We also implement these algorithms 
	employing 3DR Solo drone and ten \gds, the first such realistic testbed built.
\end{itemize}
}

\vspace{5pt}
The rest of the paper is organized as follows.
Section~\ref{sec:related} reviews the existing literature on localization approaches relevant to our context.
Section~\ref{sec:mes-er} derives expressions to approximate the measurement and ground errors,
\revision{and introduces how our results are interpreted in the light of A2G communications}.
Section~\ref{sec:loc-error} investigates the localization error
affecting the estimated position of the \gd when the trilateration procedure is applied,
and describes two localization algorithms based on trilateration compared in Section~\ref{sec:ev}.
Section~\ref{sec:algorithms} introduces four more localization algorithms not based on trilateration which are also compared using the testbed. Three of them are transformed from range-free to range-based algorithms.
Section~\ref{sec:ev} presents
a rich set of real experiments on the field aiming to evaluate the localization error
of the different localization algorithms.
Finally, Section~\ref{sec:concl} offers conclusions with directions of future research.

\vspace{-0.1in}
\section{Related Works}\label{sec:related}
\revision{This section reviews the relevant literature on localization of \gds using \mas and also efforts on testbed implementations considering UAVs as \mas.}

\vspace{-0.1in}
\subsection{\revision{\ma-based Localization Algorithms}}
There exist many algorithms for ground \mas 
that can be  
classified as range-free and range-based.
\revision{In the range-free localization algorithms, such as 
\xiao~\cite{xiao2008distributed} (intersection of circles) and 
\lee~\cite{lee2009localization} (intersection of annuli),
a rover broadcasts its current position at regular time intervals while following a path. 
From the heard and not-heard rover's positions (informally, HnH technique), the \gd builds a limited area where it may reside and places itself at the ``center''.
(More details about these algorithms are in Section~\ref{sec:algorithms}.)}

Usually range-free algorithms have relatively low localization \revision{accuracy}, 
leading to the development of range-based algorithms 
as like \scan and \dscan~\cite{koutsonikolas2007path}.
In \scan and \dscan, the \ma follows a path formed by vertical straight lines interconnected by horizontal lines.
However, such algorithms result in a large number of collinear anchor points.
Collinearity can be reduced by increasing the changes of direction in the path, 
as in \hilbert~\cite{koutsonikolas2007path} and
\lmat~\cite{jiang2011lmat}.
The path generated by \lmat logically tessellates the deployment area by equilateral triangles so that
each \gd falls inside a triangle.
The vertices of the triangle where the \gd resides are used to trilaterate the \gd position,
thus completely solving the collinearity issue.
The above algorithms are designed for ground \mas.

\revision{A few localization algorithms have  been proposed for flying \mas.
Since the drone flies at a certain altitude, there are new constraints on the anchor points.
In~\cite{perazzo2017drone}, one can find simulation comparisons of the above algorithms
extended to flying \mas, with particular attention to the path length.}
\revision{The \omni~\cite{bettisorbelli2018range} algorithm is the first localization
algorithm for drones that selects the anchor points
in such a way that  a certain accuracy 
is guaranteed.}
A simple and lightweight range-based algorithm called \drbc~\cite{bettisorbelli2019ground}
localizes the \gds by determining the correct intersection point of two circles 
exploiting a third reference point for the disambiguation. Finally, the \drf~\cite{bettisorbelli2019rangefree} algorithm,
the first range-free algorithm for UAVs as \mas,
exploits lightweight geometrical rules for position estimation.
\revision{(See Section~\ref{sec:algorithms} for more details on \omni and \drf.)}

\revision{Very recently, in~\cite{ebrahimi2020autonomous}, a novel framework
based on reinforcement learning (RL) has been proposed to enable a UAV to autonomously find a suitable trajectory. This improves the localization accuracy of multiple \gds minimizing the 
flight time, path length, and UAV's energy consumption.
As usual for RL techniques, an initial step is required for allowing the UAV
to be trained for an online real scenario.
However, this work does not mention how long this training phase lasts.}

\vspace{-0.1in}
\subsection{\revision{UAV based Testbeds for Localization Experiments}}
Recently, we performed preliminary experiments evaluating the range-free, radius-free \drf~\cite{bettisorbelli2019rangefree} algorithm using a UAV as \ma and relatively inexpensive antennas~\cite{bettisorbelli2019ground}. 
Our conclusion was that the performance of \drf heavily
depends on the shape of the antenna radiation pattern; precisely,
the more the omnidirectionality of the antenna, the higher is the localization \revision{accuracy}.
Employing the \dw MDEK1001 kit,
in~\cite{bettisorbelli2020rangefree},
it has been experimentally shown that other range-free radius-based algorithms like 
\xiao~\cite{xiao2008distributed} and \lee~\cite{lee2009localization}, also exhibit higher accuracy with the omnidirectionality of antennas.

\revision{
%
A real outdoor implementation of 
a range-based localization algorithm
is presented in~\cite{greco2015localization}, which aims to localize 
a radio-frequency identification (RFID) ground tag with the help of a UAV.
Initially, the RFID tag is detected by the UAV using its RFID reader. 
Then, the UAV takes hundreds of RSSI measurements and estimates the tag's position using a multilateration procedure.
Experimental results show an average localization error of $6 \unit{m}$.
Note that this algorithm considers only random paths and random measurements, 
and does not investigate the relationship between the localization error and the UAV's altitude.

A range-based algorithm is experimentally tested in~\cite{grigulo2018experimenting}, in which a UAV regularly broadcasts its current GPS position while flying.
The \gd aims to detect a set of at least three equidistant RSSI distance measurements from its center
in order to apply trilateration.
Experimental results show an average error of $4 \unit{m}$ under normal conditions that
reduces to $1 \unit{m}$ using GPS corrections provided by a Real Time Kinematic (RTK) algorithm.
However, an RTK positioning system is not always available.

In~\cite{cisek2017ultra} a testbed deploying five UWB ground anchors is implemented for evaluating and tracking the 3D position of a UAV equipped with UWB antenna. 
The anchors and the UAV are also equipped with GPS receivers with RTK capabilities.
The experimental error between UWB and RTK distance measurements ranges from $2$--$24 \unit{cm}$,
while the GPS positioning error alone is $2 \unit{m}$ on an average.
Another similar set-up employing UWB technology
is proposed in~\cite{lazzari2017numerical} to localize a moving UAV.
In this scenario, four fixed UWB anchor devices are placed on the ground.
Experimental results show an average localization error of $1 \unit{m}$.
Notice that, differently from the setting considered in this paper, the last two approaches~\cite{cisek2017ultra, lazzari2017numerical} aim
at localizing or tracking the UAV (\ma) instead of the \gds.
Nonetheless, they employ UWB and UAV technologies.
}

\vspace{-0.1in}
\section{Measurements and Ground Errors}\label{sec:mes-er}
In this section, we provide analytical 
bounds on the impact of different measurement errors that may occur on the estimated ground distance between 
the \ma and a \gd. 

\vspace{-0.05in}
\subsection{Terminology and Notations}
\begin{figure}[htbp]
	\centering
	\def\svgscale{1.0}
	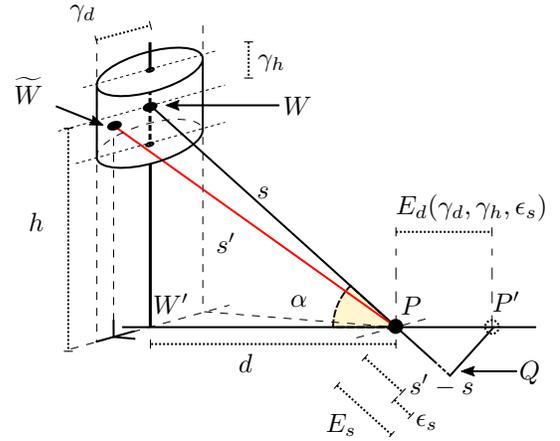
	\caption{The ground error: the point $P$ is the \gd.}
	\label{fig:ground_error_combined}
\end{figure}

Let the {\em slant distance} $s$ denote the 3D distance between the drone as \ma and the \gd. 
We define the {\em \revision{accuracy}} as the maximum error in absolute value, and we
let $\epsilon_s$  denote the {\em instrumental \revision{accuracy}}, i.e., the maximum error in estimating the slant distance.
Let the point $P$ be the \gd's position on the ground,
the point $\widetilde{W}$ be the actual drone's position, and
the point $W$ be the scheduled drone's position (see Figure~\ref{fig:ground_error_combined}).
The measured slant distance $s'= \overline{\widetilde{W}P}$ can be different 
from the exact slant distance $s=\overline{WP}$
due to the instrumental error and due to the 
\revision{accuracy} of the drone's position (i.e., the drone resides  at $\widetilde{W}$ and not at the scheduled position $W$).
Let the {\em slant error} $E_s$  be the 3D measurement error that affects the measured slant distance $s$.
The slant and the instrumental errors are depicted along with $s$ in Figure~\ref{fig:ground_error_combined}.

\vspace{-0.05in}

In reality, the \revision{accuracy} of the drone's position $\widetilde{W}$ depends on its drift
with respect to its trajectory, and on the changes of its altitude.
Indeed, a drone is more unstable than a rover, even if it hovers in a given position. 
We say that the drone {\em rolls} when it drifts to some directions on a fixed plane, and that
the drone {\em uplifts} or {\em downfalls} when 
it elevates or decreases its altitude, respectively.
We denote with $\gamma_d$ and $\gamma_h$, respectively, the {\em rolling} and the {\em altitude \revision{accuracy}} 
that depend on the GPS, and the barometer.
Figure~\ref{fig:ground_error_combined} depicts the cylinder where the drones may 
reside due to the rolling and altitude errors. 
Interestingly, the drone resides inside a cylinder instead of a sphere since we 
consider $\gamma_d$ and $\gamma_h$ independently to each other.
Let $\alpha$ in Figure~\ref{fig:ground_error_combined} be the elevation angle as when there are
no errors. That is, $\alpha$ is given assuming the scheduled position of the drone.

To localize the \gd, we must convert  the 3D slant distance $s$ 
into the  {\em ground distance} $d$, which is a distance derived on the 2D plane.
The exact ground distance $d$ is the 
distance $\overline{W'P}$ between $P$ and the projection $W'$ on the ground
of the  drone's position $W$.
That is, $d$ assumes the drone to be in the scheduled position $W$.
However, we do not know $s$, but we know $s'$.
Then, let the {\em ground error} $E_d$ be the  measurement error $\overline{PP'}$, where
$P'$ is the position of $P$ estimated on the ground by using the measured slant distance $s'$ and the scheduled elevation angle $\alpha$ which assumes the exact ground distance $d$ and the scheduled altitude $h$.
Finally, let the {\em ground \revision{accuracy}} $\epsilon_d$ be the maximum $E_d$.

\begin{table}[htbp]
\vspace{-0.1in}
	\renewcommand{\arraystretch}{1.15}
	\caption{Summary of Notations for errors and \revision{accuracies}.}
	\label{tab:nomencltature}
	\vspace{-0.1in}
	\centering
	\begin{tabular}{cl}
		\hline
		symbol & description \\
		\hline
	    $\alpha$ & elevation angle  \\
		$\epsilon_s$ & instrumental \revision{accuracy}  \\
		$\gamma_d$ & rolling \revision{accuracy}  \\
		$\gamma_h$ & altitude \revision{accuracy}  \\
		$\epsilon_d$ & ground \revision{accuracy} (max error)  \\
		$E_d$ & ground error  \\
		$E_s$ & slant error  \\
		$E_L$ & localization error  \\
		$E^T_L$ & localization trilateration error  \\
		$\epsilon^T_L$ & localization trilateration \revision{accuracy} (max error)  \\
		\hline
	\end{tabular}
\end{table}

The ground error $E_d$ is the 3D slant error $E_s$ as it is perceived on the ground.
With a single measurement, we only know the relative distance between the \ma and the \gd, thus the \gd is not yet localized.
Beyond the ground error, there is the {\em localization error} $E_L$, 
which is instead the distance from the \gd's estimated position (by any localization algorithm)
and the \gd's actual position.
This error also depends on the invoked algorithm and its implicit rules to find the \gd's position, 
and will be investigated farther. 

Table~\ref{tab:nomencltature} summarizes the notations used in this paper. 

\vspace{-0.05in}
\subsection{The Ground Error}
In this section, 
we analytically study the ground error
$E_d$
by breaking it up 
into  three independent components
$E_d(\epsilon_s)$, $E_d(\gamma_d)$, and  $E_d(\gamma_h)$
that, respectively, depend on:
\begin{inparaenum}
	\item the {\em instrumental} \revision{accuracy}, 
	\item the {\em rolling} \revision{accuracy}, and 
	\item the {\em altitude} \revision{accuracy}. 
\end{inparaenum}
We recall that we define {\em \revision{accuracy}} as the maximum error in absolute value. 
$E_d(\gamma_d)$ and  $E_d(\gamma_h)$ model the error in the drone's position.
Note that each component depends on an independent hardware part, namely, UWB, GPS, and barometer, 
and thus it makes sense to study them separately.
Whenever we study one component, we assume the other errors to be null.

\vspace{-0.05in}
\subsubsection{Instrumental error}\label{ss:ic}
\begin{figure}[htbp]
    \centering
	\def\svgscale{0.8}
	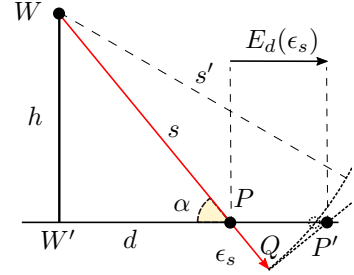
	\vspace{-0.05in}
	\caption{Overestimation in the instrumental \revision{accuracy} $\epsilon_s$.}
	\label{fig:ranging_precision}
	\vspace{-0.05in}
\end{figure}

Let us investigate $E_d(\epsilon_s)$, i.e.,
the impact of the {\em instrumental error} $e_s$ on $E_d$. 
Note that $e_s$ is defined as the difference, positive (overestimation) or negative (underestimation), 
between the measured distance and the actual distance. 
Moreover,  $\epsilon_s$ is the absolute value of the maximum instrumental error. 
Accordingly, $|e_s| \le \epsilon_s$.
Here we assume $\gamma_d=\gamma_h=0$.
Let $s$ be the exact 3D distance between the drone and the object $P$ ($P$ denotes the \gd's position).
Then, let $s'=s+e_s$ be the measure of the segment $\overline{WP}$, where $-\epsilon_s \le e_s \le \epsilon_s$.
In the following, we geometrically show how the measured slant distance $s'$ is converted into the ground distance. 
Figure~\ref{fig:ranging_precision} illustrates the
reasoning behind the choice of $e_s= \pm \epsilon_s$.
We draw a circumference of radius $s'$ centered at the waypoint $W$.
Such circumference will intersect
the line that passes between $W$ and $P$ in $Q$ (see Figure~\ref{fig:ranging_precision}).
Since the measured slant distance is different from the exact one, i.e., 
$s' \not = s$, 
$Q$ does not coincide with $P$, and $Q$ is not at the ground level.
Specifically, the segment $\overline{PQ}$ of length $e_s=s'-s$
is on the extension of $\overline{WP}$ if $e_s >0$; whereas $\overline{PQ}$ is on the radius $\overline{WP}$ if $e_s < 0$.

Since in general $\epsilon_s \ll s$, we can approximate the circumference of radius $s'$ with
its tangent in $Q$. 
The point $P'$, where the tangent intersects the 
ground\footnote{Figure~\ref{fig:ranging_precision} shows the intersection $P'$ between the tangent and the ground, which
approximates the intersection (white dot) between the circumference and the ground.
However, the two intersections become closer and closer when $s$ increases.}, is 
the estimated position for $P$ according
to the measurement $s'$.
Thus, recalling that $W'$ is the projection of $W$ on the ground, 
$\overline{PP'}$ is the error on the ground derived from the slant error $e_s$.
By elementary geometric rules applied to the right-angled triangle $PQP'$,
we obtain
$\overline{PP'} = 
e_s \cdot \frac{1}{\cos(\alpha)} = 
e_s \cdot \sqrt{1 + \frac{h^2}{d^2}}$, where $h$ is the drone's altitude.
because $\angle{QPP'}$ is equal to the elevation angle $\alpha$.
The error $E_d (\epsilon_s)$, when the instrumental error is maximum 
and the object is at ground distance $d$ from the drone, is given by:
\vspace{-0.05in}
\begin{equation}\label{eq:ground_distance_instrumental}
    E_d(\epsilon_s) = 
        \epsilon_s \cdot \frac{1}{\cos(\alpha)} = 
        \epsilon_s \cdot \sqrt{1 + \frac{h^2}{d^2}}.
\end{equation}

The ground error $E_d(\epsilon_s)$ 
varies with the distance $d$ on the ground.
When $h \not = 0$, the error increases when $d$ decreases (whereas, when $h=0$ the error does not depend on $d$). 
When $h \not = 0$, the worst case occurs when
the drone is perpendicular to the point to be measured (i.e., $W'=P$, $d=0$, $E_d \rightarrow \infty$).
From this observation, we can assert that, when the measurements are taken by a UAV, 
rather than a rover, 
in order to bound $E_d(\epsilon_s)$, 
it is convenient to add the constraint that all the measurements have 
to respect a given {\em minimum ground distance} \dmin. 

\vspace{-0.05in}
\subsubsection{Rolling error}
\begin{figure}[htbp]
	\centering
	\def\svgscale{0.8}
	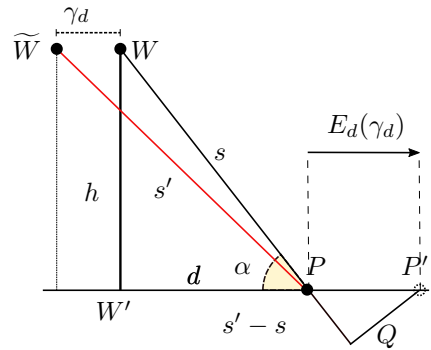
	\vspace{-0.05in}
	\caption{The rolling \revision{accuracy} $\gamma_d>0$ and ground error.}
	\label{fig:ranging_precision_gamma_d}
\end{figure}

In this section, we only consider the rolling error (i.e., $\epsilon_s=\gamma_h=0$).
When the drone hovers in position $W=(x,y,z)$, 
it may not be in  $W$, but rather
in position $\widetilde{W}$ due to the GPS \revision{accuracy} or the bad weather conditions (see Figure~\ref{fig:ranging_precision_gamma_d}). 

To better define the rolling error, we set a 3D-Cartesian coordinate system whose origin is the projection $W'=(0,0,0)$ on the ground of 
the exact drone's position $W$, whose $x$-axis passes through the object to measure $P$, and $z$-axis passes through $W$. 
Thus, $W=(0,0,h)$ and $P=(d,0,0)$. 
Then, let the actual drone's position be $\widetilde{W}=(e_x,e_y,h)$, with $-\gamma_d \le e_x, e_y \le \gamma_d$, where $\gamma_d$ is the rolling \revision{accuracy}.
Obviously, $\widetilde{W}'=(e_x,e_y,0)$ is the projection of $\widetilde{W}$ on the ground, which is inside a circle of radius $\gamma_d$ centered at the origin $W'$.
For each point of the circle, it holds  $e_x = \gamma_d \cos(\psi)$ and $e_y = \gamma_d \sin(\psi)$, where 
$\psi=\angle{\widetilde{W}'W'P}$ and $0 \le \psi \le 2\pi$.
The measured slant distance $s'$ between $\widetilde{W}$ and $P$  given by:
\begin{align*}
    s'&=\sqrt{h^2 + (d-e_x)^2 + e_y^2}  \nonumber \\
    &=\sqrt{h^2 + (d-\gamma_d \cos(\psi))^2 + (\gamma_d \sin(\psi))^2} \nonumber \\
    &= \sqrt{h^2 + d^2 - 2d\gamma_d\cos(\psi) + \gamma_d^2 \cos^2(\psi) + \gamma_d^2 \sin^2(\psi) } \nonumber \\
    &= \sqrt{h^2 + d^2 - 2d\gamma_d\cos(\psi) + \gamma_d^2 }\nonumber
\end{align*}
Recalling that $h>0$, $d>0$, $\gamma_d \geq 0$ and $0 \le \psi \le 2\pi$, we note that $s'$ is maximum when 
$f = - 2d\gamma_d\cos(\psi)$ is minimum. 

Including the instrumental error, 
the slant error is:
$$E_s = s'-s= \sqrt{h^2 + d^2 - 2d\gamma_d\cos(\psi) + \gamma_d^2}-\sqrt{h^2+d^2}$$
which is 
maximum when $\cos(\psi)=-1$ and $\epsilon_s > 0$.

In order to project $E_s$ on the ground, we repeat the same construction
as in Section~\ref{ss:ic}.
We draw a circumference 
of radius $s'$ centered in the waypoint $W$, which intersects
the line that passes for $W$ and $P$ in $Q$. 
The tangent in $Q$ intersects the ground in the estimated position $P'$.
Applying elementary trigonometry to the right-angled triangle $PQP'$
whose $\angle{QPP'}$ is equal to the elevation angle $\alpha$, 
\begin{align}
    E_d(\gamma_d) &= \frac{E_s(\gamma_d)}{\cos(\alpha)} 
    = \frac{\left| s'- s \right|}{\cos(\alpha)}
    = \frac{|(s')^2 - (s)^2|}{\cos(\alpha)\left( s' + s \right)}  \nonumber \\
    &= \frac{|\gamma_d^2 - 2d\gamma_d \cos(\psi)|\sqrt{h^2 + d^2}}{\left( s' + s \right)d}
    = \frac{|\gamma_d^2 - 2d\gamma_d \cos(\psi)|s}{\left( s' + s \right)d}  \nonumber
\end{align}

When $s'>s$ (i.e., $\frac{\pi}{2} < \psi < \frac{3\pi}{2}$), that is, when the drone rolls away from the object, 
it holds:
\begin{align}
    \label{eq:ground_distance_gamma_d_roll_away}
    E_d(\gamma_d) &\le \frac{|\gamma_d^2 - 2d\gamma_d|}{2 d } \nonumber \\
    \intertext{and assuming $\gamma_d \ll d$}
    E_d(\gamma_d)  &\le \gamma_d 
\end{align}

When $s'< s$ (i.e., $0 \le \psi \le \frac{\pi}{2}$ or $\frac{3\pi}{2} \le \psi \le 2\pi$),
that is, when the drone rolls close to the object, since $s+s'> s$, we obtain a weaker bound:
\begin{align*}	
    E_d(\gamma_d) & < \frac{|\gamma_d^2 - 2d\gamma_d|}{d}\frac{s}{s}  \nonumber \\
    & < 2\gamma_d
\end{align*}
Now, if $\gamma_d \ll d$ holds, $\frac{s}{s'} \rightarrow 1$.
Since $s+s' \ge 2s'$, we have:
\begin{align}
    \label{eq:ground_distance_gamma_d_bound_strict}
    E_d(\gamma_d) & < \frac{|\gamma_d^2 - 2d\gamma_d|}{2d}\frac{s}{s'} \nonumber \\
    & < \gamma_d\frac{s}{ s'} \rightarrow \gamma_d 
\end{align}

We will see in our experiments that indeed the stricter bound
in Eq.~\eqref{eq:ground_distance_gamma_d_bound_strict}
well approximates the rolling error even when the drone rolls close to the \gd.

\vspace{-0.05in}
\subsubsection{Altitude error}
\begin{figure}[htbp]
	\centering
	\def\svgscale{0.8}
	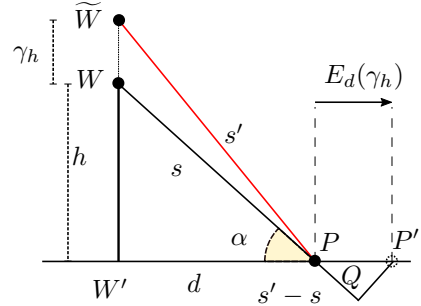
	\vspace{-0.05in}
	\caption{The altitude \revision{accuracy} $\gamma_h> 0$ and ground error.}
	\label{fig:gamma_h_error}
\end{figure}

In this section, we only consider the altitude error (i.e., $\gamma_d=\epsilon_s=0$).
When the drone is subject to an uplift (resp., downfall), 
the measured slant distance $s'$ is overestimated (resp., underestimated).
The overestimate case is illustrated in  Figure~\ref{fig:gamma_h_error}.

The measured slant distance $s'$ between $\widetilde{W}$ and $P$ is: 
$s' = \sqrt{h+{\gamma_h}^2 + d^2}$.
Recalling that $h>0$, $d>0$, and $\gamma_h \geq 0$, including the instrumental error, 
the slant error is:
$$E_s = s'-s= \sqrt{h+{\gamma_h}^2 + d^2} - \sqrt{h^2+d^2}$$
Moreover,
\vspace{-0.1in}
\begin{align}
    E_d(\gamma_h) &= \frac{E_s(\gamma_h)}{\cos(\alpha)} 
    = \frac{\left| s'- s \right|}{\cos(\alpha)} 
    = \frac{|\gamma_h^2 - 2h\gamma_h|s}{\left( s' + s \right)d} \nonumber
\end{align}

Repeating  calculations similar to those above, 
and assuming that the altitude \revision{accuracy} $\gamma_h$ is very small 
with respect to $h$, and that $\frac{s}{s'} \rightarrow 1$,
we find that the ground error can be approximated as:
\vspace{-0.1in}
\begin{equation}\label{eq:ground_distance_gamma_h_bound}
    E_d(\gamma_h) \approx  \gamma_h \frac{h}{d}
\end{equation}

\subsubsection{Overall ground error}
From the previous discussions we can estimate the overall ground error as stated by the following:

\begin{fact}
	Let $\epsilon_s$, $\gamma_d$, and $\gamma_h$  be respectively the instrumental \revision{accuracy}, rolling \revision{accuracy}, and altitude \revision{accuracy} that may affect the slant measurement.
	By projecting the slant distance on the ground, the largest error $E_d$ given the ground distance $d$, is:
	
	\vspace{-0.1in}
	\begin{equation}\label{eq:Ed}
	    E_{d}(\gamma_d, \gamma_h, \epsilon_s) \approx \gamma_d + \frac{h}{d}\gamma_h + \epsilon_s\sqrt{1+\frac{h^2}{d^2}}
	\end{equation} 
\end{fact}

Analyzing Eq.~\eqref{eq:Ed}, it is clear that when $d$ is very small, the ground error is very large.
Increasing $d$, 
the impact of both instrumental and altitude \revision{accuracies} decreases, but
$E_d$ cannot be smaller than the rolling \revision{accuracy} $\gamma_d$.

In conclusion, the ground error can be  bounded by adding a
constraint on the minimum 
ground distance (\dmin) between the drone and the \gd.
If it is ensured that $d \ge \dmin$ using the drone, then	
the {\em ground \revision{accuracy}} $\epsilon_d$, i.e., the maximum error
on the ground distance, is bounded by:
\begin{equation}\label{eq:ground_distance_combined_bound_def}
    \epsilon_d = \epsilon_d(\gamma_d, \gamma_h, \epsilon_s) \approx \gamma_d + \frac{h}{\dmin}\gamma_h + \epsilon_s\sqrt{1+\frac{h^2}{\dmin^2}}
\end{equation}

\revision{Our first takeaway is
that the ground \revision{accuracy} $\epsilon_d$ can be monitored by monitoring
the ratio  $h/d$ between altitude and ground distance.
}

\vspace{5pt}
\revision{\paragraph{A2G links and ground error}
In this paragraph, we explain how the A2G communication link quality
between \gd and \ma impacts on our results.
According to the model in~\cite{al2014optimal},
each A2G link has a certain probability $P(\text{LoS})$ to be in LoS
and $P(\text{NLoS})$ to be in NLoS.
$P(\text{LoS})$ depends on the elevation angle $\alpha$ between drone and \gd
and on the environment type, i.e., sub-urban, urban, dense, and highrise.
Clearly, in crowded environments, 
links have a higher probability to be mixed LoS and NLoS scenarios.

\begin{table}[ht]	
\vspace{-0.1in}
	\renewcommand{\arraystretch}{1.15}
	\caption{The line of sight probabilities $P(\text{LoS})$ in different environments~\cite{al2014optimal}.}
	\label{tab:lap}
	\vspace{-0.1in}
	\centering
	\begin{tabular}{cc|cccc}
		\hline
		$h/\dmin$ & $\alpha$ & sub-urban & urban & dense & highrise \\
		\hline
		$5.67$ & $80^{\circ}$ & $100\%$ & $100\%$ & $100\%$ &$100\%$ \\
		$\sqrt{3}$ & $60^{\circ}$ & $100\%$ & $100\%$ & $100\%$ &$60\%$ \\
		$1$ & $45^{\circ}$ & $100\%$ & $97\%$ & $85\%$ & $30\%$ \\
		${1}/{2}$ & $26.5^{\circ}$ & $100\%$ & $75\%$ & $30\%$ & $5\%$ \\
		${1}/{3}$ & $20^{\circ}$ & $100\%$ & $40\%$ & $20\%$ & $\to 0\%$ \\
		\hline
	\end{tabular}
	\vspace{-0.05in}
\end{table} 

The UWB distance measurements are possible as long as the antennas keep an A2G link.
Up to $35 \unit{m}$, even if the elevation angle is small,
communications can be established since UWB works in both LoS and NLoS~\cite{www-Deca-ieee}.
Beyond $35 \unit{m}$ UWB only works in LoS, and hence only LoS links can be guaranteed.
Suitable values for $h/\dmin$
such that the elevation angle $\alpha=\arctan(h/\dmin)$ 
gives LoS links with high probability, have to be selected.
For example, as reported in Table~\ref{tab:lap},
in sub-urban environment $h/\dmin=1/3$ because A2G link has $100\%$ probability to be LoS whenever $\alpha \ge 20^{\circ}$.
In urban environment, links are $100\%$ in LoS when $h/\dmin \ge 1/2$  ($\alpha \ge 26.5^{\circ}$).
Similarly, in highrise the minimum ratio for LoS links is $h/\dmin=5.67$.
Note that for $h/\dmin < 1/3$ ($\alpha < 20^{\circ}$) links could be mixed and hence UWB might work or not.
Recall that the ground accuracy in Eq.~\eqref{eq:ground_distance_combined_bound_def}
can be bounded selecting a small $h/\dmin$.
Keeping in mind the maximum UWB NLoS range of $35 \unit{m}$,
$h/\dmin$ cannot be freely chosen. 
However, in our experiments since we work in a sub-urban, obstacle-free, and flat environment,
any ratio $h/\dmin \ge 1/3$ is sufficient to be in LoS.
}

\vspace{-0.1in}
\section{Localization Error for Trilateration Based Algorithms}\label{sec:loc-error}
Once a \gd has collected a suitable number of distance measurements from the \ma,
it can be localized by invoking any localization algorithm. 
A very common approach for localization is {\em trilateration}. 
In Section~\ref{sec:error-trilateration} we analytically derive 
the {\em localization trilateration error} $E^T_L$ and
the {\em localization trilateration \revision{accuracy}} $\epsilon^T_L$,  
which are incurred by  any algorithm based on  this method. 
Subsequently, in Section~\ref{sec:omni-scan-algs}, we discuss the trilateration based algorithms that are considered in our experiments. 

\begin{figure*}[ht]
	\centering
	\subfloat[Linearization of each measurement.]{%
		\def\svgscale{0.85}
		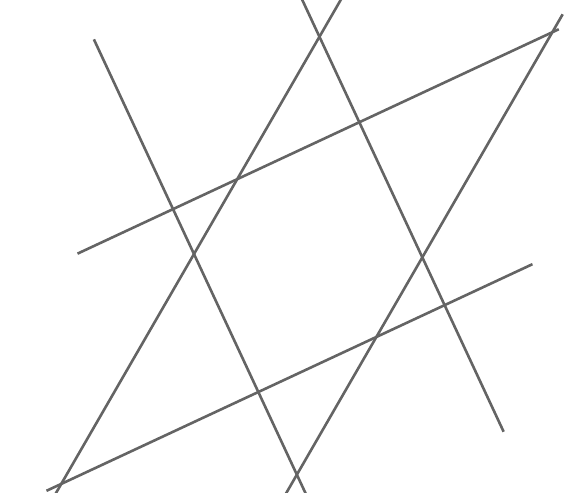
		\label{fig:trilateration_error}
	}
	\subfloat[Same signs estimation.]{%
		\def\svgscale{0.85}
		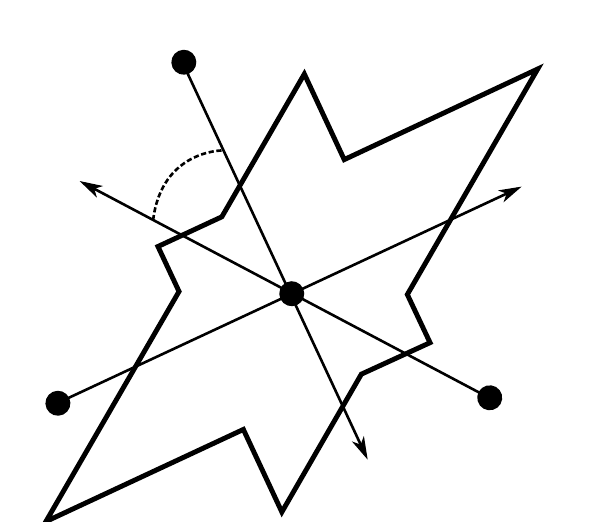
		\label{fig:trilateration_error_sin}
	}
	\subfloat[Different signs estimation.]{%
		\def\svgscale{0.85}
		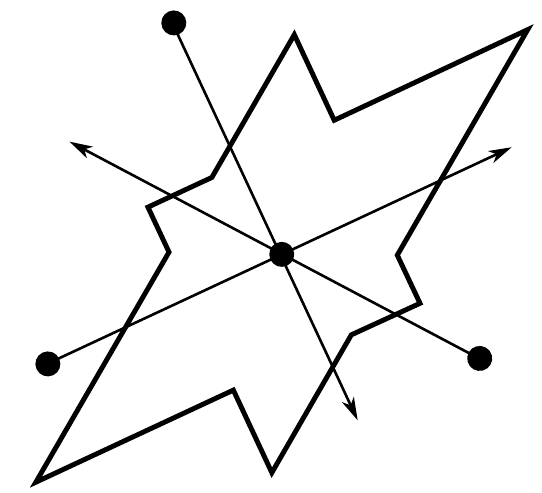
		\label{fig:trilateration_error_cos}
	}
	\caption{Trilateration error with different conditions.}
	\label{fig:trilateration_error_proof}
	\vspace{-0.1in}
\end{figure*}

\vspace{-0.05in}
\subsection{Trilateration Error}\label{sec:error-trilateration}
This section discusses the localization error $E^T_L$ that may affect the 
estimated \gd's position when the trilateration procedure
is applied. 

Let us briefly recall that the trilateration procedure 
for estimating the position of the object $P$, takes as input three
ground distances $d_1$, $d_2$, and $d_3$ of $P$ from three waypoints $W_1$, $W_2$, and $W_3$ respectively.
The procedure returns, as the  \gd's estimated position $P$, the intersection of the three circumferences
corresponding to the radii $d_1$, $d_2$, and $d_3$ centered at the projections 
$W'_1$, $W'_2$, and $W'_3$ of the waypoints.
Due to the ground errors, however, the three circumferences do not intersect
at a single point, but they delimit a small {\em star} area,
as depicted in Figure~\ref{fig:trilateration_error}. 
In fact, a pair of extreme circumferences, one obtained by considering the radius affected by the maximum positive 
$E_d$ error ($d_i + E_d$, measurement overestimation) and one whose
radius is affected by the maximum negative $E_d$ error ($d_i - E_d$, measurement underestimation)
is drawn in place of each circumference of radius $d_i$.

Assuming that all the ground distances are sufficiently large compared to the ground error,
these extreme  circumferences can be linearized (i.e., replaced by the tangent to the radius) without significantly changing  the area. 
Each different non-parallel pair of linearized circumferences 
intersects at a single point forming overall $12$ 
different points, that correspond to the vertices of the star shape.
Note that $P$ is at the center of the star.
The trilateration procedure returns as the estimated position, instead of the exact intersection $P$, a point $P'$ in the star.
The point $P'$ is selected
by means of the least-squares-error method.
In fact, given three ground measurements, the estimated position of $P$ is the point $(x_P,y_P)$ 
that minimizes the sum of the least squares, i.e.:
\vspace{-0.05in}
\begin{equation}\label{eq:trilateration}
    \begin{array}{r@{}r@{}r@{}l}
        \text{min} \quad \delta^2_1 + \delta^2_2 + \delta^2_3 \\[\jot]
        \text{s.t.}\qquad \sqrt{(x_{W'_i}-x_P)^2+(y_{W'_i}-y_P)^2} &{} +\delta_i=\overline{W'_i P}  \\
        \multicolumn{4}{c}{ \hspace{4.9cm} \textrm{for}  \quad i=1,2,3.}
    \end{array}
    \vspace{-0.05in}
\end{equation}
The largest value of the positioning error, i.e., $\overline{PP'}$, 
called {\em localization trilateration error} $E^T_L$,
or simply {\em trilateration error},
occurs when the estimated position $P'$ is at the furthest vertex of the star shape.
In other words, the positioning error is bounded by the distance between the center of the star $P$ (i.e., the actual position of the \gd)
and its farthest vertex.

\begin{figure}[htbp]
\vspace{-0.1in}
	\centering
	\subfloat[The three angles $\beta_i$.]{%
		\def\svgscale{0.65}
		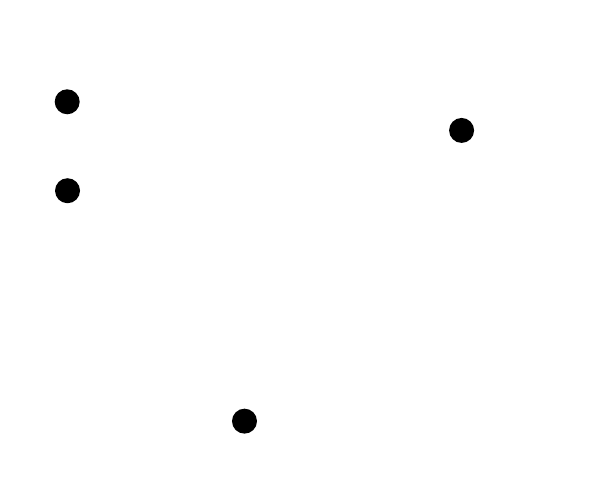
		\label{fig:angular_aperture}
	}
	\subfloat[$E^T_L$.]{%
		\includegraphics[scale=0.775]{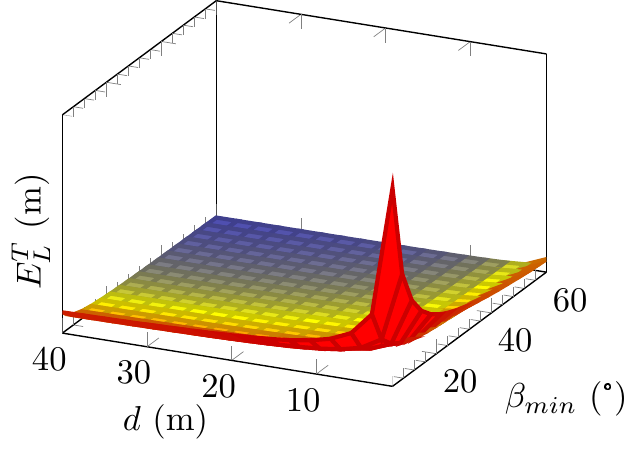}
		\label{fig:precision_vs_angle_distance}
	}
	\caption{The trilateration error $E^T_L$: when $d$ and $\bmin$ are very small, the error is extremely high.}
	\label{fig:angles}
\end{figure}

As an example, in Figure~\ref{fig:trilateration_error_sin}, 
the distance between the actual point $P$ and 
the estimated point $P'$ 
at the intersection of  two measurement underestimations  $d_2 (-)$ and $d_3 (-)$  
is $\frac{E_d}{\cos(\beta/2)}$, where $\beta$ is one of the three different
angles in which the turn angle in $P$ is divided by the lines $\overline{W'_1P}$, $\overline{W'_2P}$, and $\overline{W'_3P}$ (see Figure~\ref{fig:angular_aperture}).
In Figure~\ref{fig:trilateration_error_cos}, 
the distance between $P$ and $P'$ that results from the measurement underestimation, i.e., $d_1 (+)$,
and the measurement overestimation, i.e., $d_3 (-)$, is depicted.
In this case, the distance $\overline{PP'}=\frac{E_d}{\sin(\overline{\beta}/2)}$.

For each vertex of the star, depending on the signs of the estimations 
($+$ overestimation, $-$ underestimation) of each pair of circumferences, we have:
$\frac{E_d}{\sin(\beta_i/2)}$ if the signs are the same; and 
$\frac{E_d}{\cos(\beta_i/2)}$ if the signs are different,
where $\beta_1 \le \beta_2 \le \beta_3$ are the three different angles formed in $P$
such that   $\sum_i \beta_i = \pi$.
In the following, we  prove that the farthest vertex occurs when the measurement estimations
have the same 
signs and the angle is minimum. 
\begin{lemma}[\cite{bettisorbelli2018accuracy}]
	Let $\bmin = \min_{i}\{\beta_i\}$, 
	$\bmax = \max_{i}\{\beta_i\}$ and $\sum_i \beta_i = \pi$. 
	Then $\sin ( \frac{\bmin}{2} ) \leq \cos ( \frac{\bmax}{2} )$.
\end{lemma}
\begin{IEEEproof}
	Let $\bmin = \min_{i}\{\beta_i\}$ and 
	$\bmax = \max_{i}\{\beta_i\}$. Then,  we have:
	$\bmax \leq \pi - 2\bmin \Rightarrow \frac{\bmax}{2} \leq \frac{\pi}{2} - \bmin$
	from which $\cos ( \frac{\pi}{2} -\bmin ) \leq \cos ( \frac{\bmax}{2} )$, and thus
	$\sin(\bmin) \leq \cos ( \frac{\bmax}{2} )$.
	\noindent Since $0 \leq \bmin \leq \pi/3$, it yields:
	\begin{gather}
	    \sin \left( \frac{\bmin}{2} \right) \leq \sin(\bmin) \leq \cos \left( \frac{\bmax}{2} \right) \nonumber
	\end{gather}
	Thus, the furthest vertex is at distance $\frac{E_d}{\sin (\frac{\bmin}{2})}$ from $P$.
\end{IEEEproof}

\begin{theorem}[\cite{bettisorbelli2018accuracy}]
	Given the \revision{accuracies} $\epsilon_s$, $\gamma_d$, and $\gamma_h$, given \dmin, and recalling that
	$\epsilon_{d}(\gamma_d, $ $\gamma_h, \epsilon_s) \approx |\gamma_d| + \frac{h }{\dmin}|\gamma_h| + |\epsilon_s|\sqrt{1+\frac{h^2}{\dmin^2}}$, the localization trilateration \revision{accuracy} defined as the
	maximum trilateration error is obtained as:
	\begin{equation}
    	\label{eq:eps_l}
    	\epsilon^T_L(\gamma_d, \gamma_h, \epsilon_s) = \frac{\epsilon_d(\gamma_d, \gamma_h, \epsilon_s) }{\sin \left(\frac{\bmin}{2}\right)}
	\end{equation} 
\end{theorem}
Therefore, from Eq.~\eqref{eq:eps_l}, we learn that, given a certain ground error, the localization error is minimized when
$\bmin \rightarrow \frac{\pi}{3}=60^{\circ}$.
Figure~\ref{fig:precision_vs_angle_distance} reports an example of the 
trilateration error ${E^T_L}$ computed by 
varying both the values $d$ and $\bmin$, 
and assuming only the instrumental error, i.e., 
$\gamma_d=\gamma_h=0 \unit{m}$ and $\epsilon_s = 0.10 \unit{m}$.
As expected, when both $d$ and $\bmin$ tend to $0$, ${E^T_L}$ grows quickly.

\revision{Analyzing Eq.~\eqref{eq:eps_l} it is clear
that the localization trilateration \revision{accuracy} $\epsilon^T_L$ 
can be bounded keeping the minimum angle
$\bmin$ as large as possible, i.e., closer to $60^{\circ}$.

Our second takeaway is that a good localization accuracy in trilateration methods can be obtained keeping as low as possible the ratio $h/\dmin$, the elevation angle $\alpha$ as small as possible respecting also the communication conditions in LoS,
and making the minimum angle \bmin as large as $60^{\circ}$.}

\vspace{-0.05in}
\subsection{\omni and \scan Localization Algorithms}\label{sec:omni-scan-algs}
In this section, we  review trilateration based   algorithms  \omni and \scan  considered in our experiments. Based on the above discussion, the localization error for these algorithms is bounded by  $\epsilon^T_L$, which is a  
function of the \revision{accuracies} $\epsilon_s$, $\gamma_d$, and $\gamma_h$, 
the minimum angle \bmin, altitude $h$, and minimum distance \dmin. 
Both algorithms are based on a static path \stp formed by 
a series of vertical lines (each called as {\em vertical scan}) 
connected by horizontal lines.

\begin{figure*}[htbp]
	\centering
	\small
	\subfloat[\drf.]{%
		\def\svgscale{0.50}
		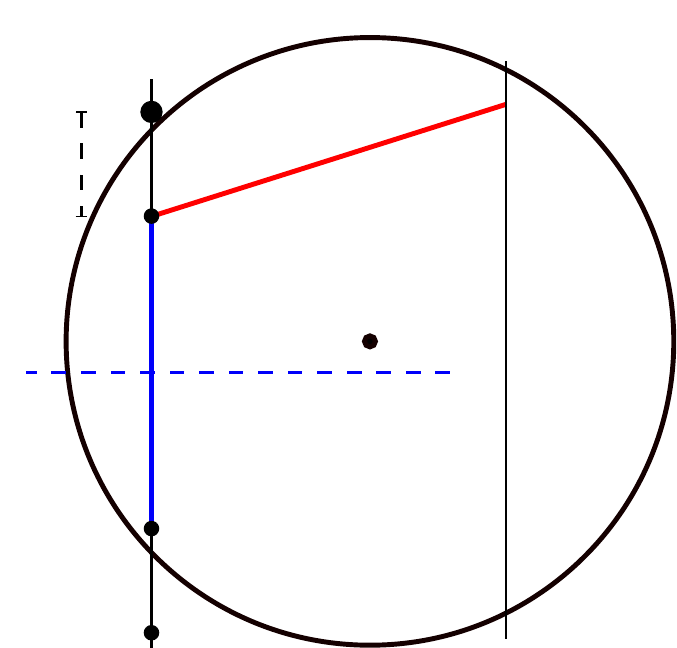
		\label{fig:model-drf}
	}
	\subfloat[\xiao.]{%
		\def\svgscale{0.50}
		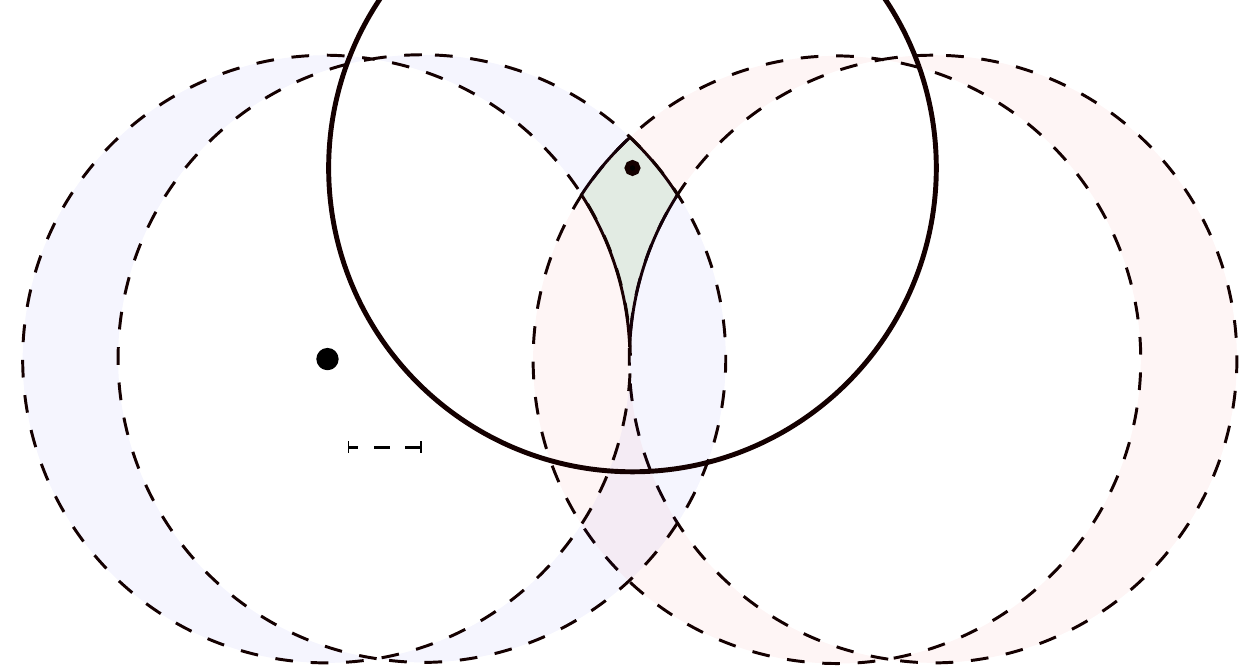
		\label{fig:model-xiao}
	}
	\subfloat[\lee.]{%
		\def\svgscale{0.50}
		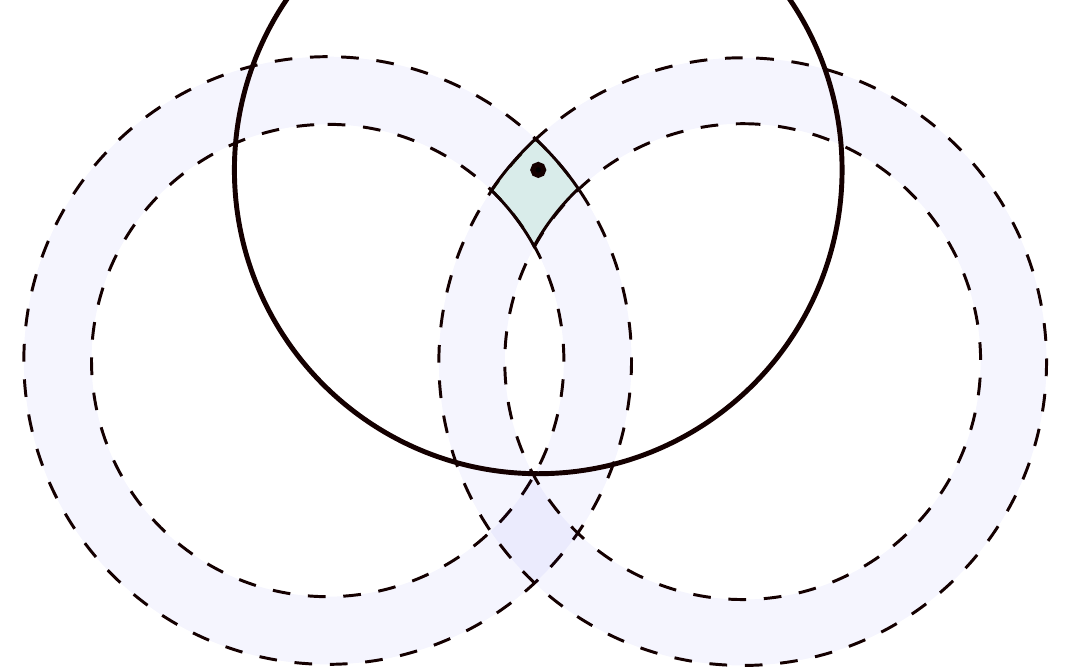
		\label{fig:model-lee}
	}
	\caption{The \drf, \xiao, and \lee localization algorithms. 
		In \xiao e \lee there are two symmetric intersection areas: a third point
		is required to find and disambiguate the correct intersection area.}
	\label{fig:intersection_areas}
	\vspace{-0.1in}
\end{figure*}

\paragraph{The \scan Algorithm}
\scan~\cite{koutsonikolas2007path} is
one of the first range-based localization algorithms designed for rovers.
Each \gd is localized employing trilateration using three waypoints.
The main drawback is the collinearity between points in the estimation phase.
Since we wish to avoid such undesirable conditions, in our experiments
we perform single trilateration selecting three non-collinear waypoints
from at least two distinct vertical scans.
In this slightly improved version of \scan,
the \bmin and \dmin constraints may not be satisfied, resulting in large  localization errors.

\paragraph{The \omni Algorithm}
\omni~\cite{bettisorbelli2018range} is 
the first range-based localization algorithm that takes into account the impact of the drone's altitude on
the measurement \revision{accuracy} and on the geometry of the waypoints from which trilateration is performed.
It logically tessellates the deployment area into  diamonds.
Then, each \gd, once it has acquired a sufficient number of waypoints/distances from the drone,
performs two trilaterations.
The first trilateration is invoked using any three non-collinear waypoints in order to compute the
logical diamond in which the \gd resides.
Since each diamond is associated with an optimal triple of waypoints 
which satisfy the minimum angle/distance constraints~\cite{bettisorbelli2018range}, 
any \gd belonging to such diamond can be finally trilaterated 
for a second time using that triple.
In conclusion, \omni has been proved to be highly accurate but requires two trilaterations, as opposed to the single trilateration performed by  \scan.

\vspace{-0.1in}
\section{Other Localization Algorithms}\label{sec:algorithms}

In this section, we describe  four more localization algorithms, namely \drf, \lee, \xiao, and \drbc, not based on trilateration. The first three of such algorithms are  {\em range-free}, therefore the localization \revision{accuracy} 
depends on the antenna radiation pattern quality.
Recently, Betti et al.~\cite{bettisorbelli2019ground}  experimentally showed  the poor \revision{accuracy} of \drf using relatively inexpensive hardware.
In this paper, motivated by these results, we extend these algorithms by considering distance measurements to improve on the localization \revision{accuracy}. 
Specifically, as also detailed in the experiments in Section~\ref{sec:ev},  
the \gd stores, for each waypoint that it hears,
the relative distance between itself and that waypoint. 
Exploiting this information, we reformulate all the range-free techniques making them actually range-based. This way, we mean to overcome the poor localization \revision{accuracy} resulting from the low-quality of the radio antenna, while  still  keeping the original procedures for the localization.

In the following, for each of these extended algorithms, as well as for \drbc, we identify  the sources of the localization error.
However, we do not derive any analytical expression of $E_L$, since the analysis would involve too many variables to be expressed in a closed formula. Nevertheless, we study their error through real experiments in  Section~\ref{sec:ev}. 

\paragraph{The \drf Algorithm}
\drf~\cite{bettisorbelli2019rangefree} is a lightweight range-free radius-free algorithm designed for drones.
This algorithm is based on the notion of {\em chord}.
In general, the perpendicular bisector of any chord passes through the center $O$ of the circle itself.
So, the bisector of another non-parallel chord and the previous one 
intersect at $O$ point.
In Figure~\ref{fig:model-drf}, two chords are identified by the pairs $A_1A_2$ and $A_2B_1$.
Theoretically, the circle is identified by the receiving disk of \gd which is centered at $O$.
Accordingly, the \gd starts to estimate its position when it detects two chords.
The two chords are detected using the HnH technique~\cite{bettisorbelli2020rangefree} on each scan.
The detection of chords incurs several problems that eventually affects the localization accuracy.
First, recalling that
the \ma regularly broadcasts its current position (waypoint) at discrete intervals of time
and that two consecutive waypoints are at distance {\em inter-waypoint} \iw,
the endpoints of the chords may not exactly fall on the circumference of the receiving disk,
even if the receiving disk is a perfect circle (e.g., $A_2$ and $A_3$ in Figure~\ref{fig:model-drf}).
However, the chords can be improperly defined if the antenna pattern has ``holes'' and  ``bubbles'',
as experienced in the field as reported in~\cite{bettisorbelli2020rangefree}.

\underline{{\em Range-based extension}:} 
Exploiting the fact that our tested kit allows us to take distance measurements,
the choice of the chords can be performed selecting three waypoints at a certain fixed distance $d$ from the \gd, 
relaxing the range-free constraint.
In this way, with three waypoints on the same circumference of radius $d$,
two chords can be derived.
Accordingly, we can obtain a localization error which depends only on
the length of \iw and on the error $E_d$.
A more detailed explanation of  the original version of \drf can be found in~\cite{bettisorbelli2019ground}.

\paragraph{The \xiao Algorithm}
\xiao~\cite{xiao2008distributed} 
is a range-free radius-based localization algorithm initially developed for ground \mas.
Like \drf, the \xiao algorithm exploits the HnH method in order to detect special points
used for building a constraint area that bounds the \gd's position.
However, diversely from \drf, \xiao relies also on the value of the communication radius $d$.
In fact, initially the \gd detects the 
pair of endpoints ($A_1$ and $A_2$, in Figure~\ref{fig:model-xiao}) using the HnH method.
Successively, two more points called pre-arrival and post-departure ($A_0$ and $A_3$),
respectively, are determined using the value of \iw, since \ma sends its current position
at discrete intervals.
Note that such four points belong on the same straight line.
Then, four circles of radius $d$ centered at each of these four points are drawn.
Those circles create two symmetrical intersection areas where the \gd may reside.
In order to select the correct intersection area, the \gd needs to detect a third point. 
Finally, the \gd is localized at the ``center'' of the correct intersection area.
This definition of center slightly varies depending on the shape of the intersection area,
which may vary from four to five vertices.

\underline{{\em Range-based extension}:} As for \drf, also in \xiao we exploit the distance measurements for computing all the required points. 
That is, we select the two waypoints on the same line at distance $d$ from \gd as $A_1$ and $A_2$, 
and the  preceding and subsequent waypoints as $A_0$ and $A_3$.

\paragraph{The \lee Algorithm}
\lee~\cite{lee2009localization},
is a range-free radius-based algorithm very similar to \xiao.
Indeed, it builds a similar constrained area using the HnH method
and the knowledge of both $d$ and \iw.
Once the \gd has detected the two extreme endpoints ($A_1$ and $A_2$, in Figure~\ref{fig:model-lee}),
it traces two circles of radius $d$ and $d-\iw$ on both the points.
These circles, which create two annuli, intersect in two distinct and symmetrical intersection areas,
so also in this case a third point is required.
Finally, the \gd estimates its position at the center of such an area, 
using easy geometrical rules.

\underline{{\em Range-based extension}:} As for the previous algorithms, we select the two extreme endpoints $A_1$ and $A_2$ 
as two waypoints at distance $d$ from the \gd.

\vspace{-0.05in}
\paragraph{The \drbc Algorithm}
\drbc~\cite{bettisorbelli2019ground} is a lightweight range-based technique designed for UAVs.
The goal of \gd is to detect two waypoints 
at distance $d_1$ and $d_2$,
and drawing two circumferences
centered at these waypoints, of radius $d_1$ and $d_2$, respectively.
Then, the \gd knows to reside simultaneously on the two intersections of two circumferences,
and a third point is required to disambiguate the correct intersection point.
In conclusion, we can note that,
differently from the trilateration based algorithms \omni and \scan
in which the least-squares-error method is employed (see Eq.~\eqref{eq:trilateration}),
\drbc only demands a few algebraic calculations.

Finally, Table~\ref{tab:algorithms_evaluation} summarizes the 
six algorithms that will be compared in our testbed.

\begin{table}[ht]
\vspace{-0.1in}
	\renewcommand{\arraystretch}{1.15}
	\caption{Summary of the compared algorithms.}
	\label{tab:algorithms_evaluation}
	\vspace{-0.05in}
	\centering
	\begin{tabular}{llcl}
		\hline
		name & method & points & error source \\
		\hline
		\omni~\cite{bettisorbelli2018range} & trilaterations & $3+3$ & geometry \\
		\scan~\cite{koutsonikolas2007path} & trilateration & $3$ & geometry \\
		\drbc~\cite{bettisorbelli2019ground} & circles intersection & $2+1$ & center \\
		\drf~\cite{bettisorbelli2019rangefree} & bisector intersection & $3$ & chords \\
		\xiao~\cite{xiao2008distributed} & points ``center'' & $2+1$ & center  \\
		\lee~\cite{lee2009localization} & points ``center'' & $2+1$ & center  \\
		\hline
	\end{tabular}
	\vspace{-0.1in}
\end{table}

\vspace{-0.1in}
\section{Evaluation on a Real Testbed}\label{sec:ev}
In this section we propose our experimental evaluation.
\revision{Initially, in Section~\ref{sec:uwb-performance} we describe the adopted hardware for our testbed.}
In Section~\ref{sec:1st-experiments}, we study the ground error $E_d$.
In Section~\ref{sec:2nd-experiments},
we study the localization error ${E^T_L}$ of the  trilateration method. 
Finally, in Section~\ref{sec:3rd-experiments},
we run a campaign of experiments with the goal of comparing
the localization error of different algorithms.

\vspace{-0.05in}
\subsection{\revision{Performance of UWB Antennas}}\label{sec:uwb-performance}
The experiments in Sections~\ref{sec:1st-experiments} and~\ref{sec:2nd-experiments}
are done using the \dw EVK1000 kit (see Figure~\ref{img:evk1000}),
formed by two UWB antennas which are based on the DW1000 UWB chip~\cite{www-Deca-dwm1000}.
For the experiments done in Section~\ref{sec:3rd-experiments}, 
we rely on a more consistent set of twelve antennas,
using the  MDEK1001 kit (see Figure~\ref{img:dwm1001}), 
based on the same DW1000 UWB chip~\cite{www-Deca-dwm1001}.

\revision{
According to \dw,
those chips have a $6.5 \unit{GHz}$ center frequency,
and have a declared and reliable point-to-point range up to $60 \unit{m}$ LoS
and $35 \unit{m}$ NLoS~\cite{www-Deca-dwm1001}
on a typical use-case. 
Although the DW1000 chip transmitting power is set to $-41.3 \unit{dBm/MHz}$,
and the typical receiver sensitivity is $-93 \unit{dBm/500 MHz}$~\cite{www-Deca-dwm1000},
the received power is influenced by the antenna polarity. 
In both the \dw kits the antennas are vertically polarized, 
meaning that the module is intended to be vertically positioned to 
let another vertically polarized antenna observe
an omnidirectional radiation pattern in the azimuth plane~\cite{www-Deca-dwm1001}.
For this reason, following the recommendations provided by \dw in their datasheet,
in our experiments we always vertically placed our antennas.
The antenna placed on the drone is positioned vertically, but reversely,
keeping the transceiver on the bottom, for avoiding the \ma's body to become an obstacle between the \gds and the \ma itself.

\begin{figure}[htbp]
\vspace{-0.2in}
	\centering
	\hfill
	\subfloat[EVK1000 kit.]{%
		\includegraphics[height=3.0cm]{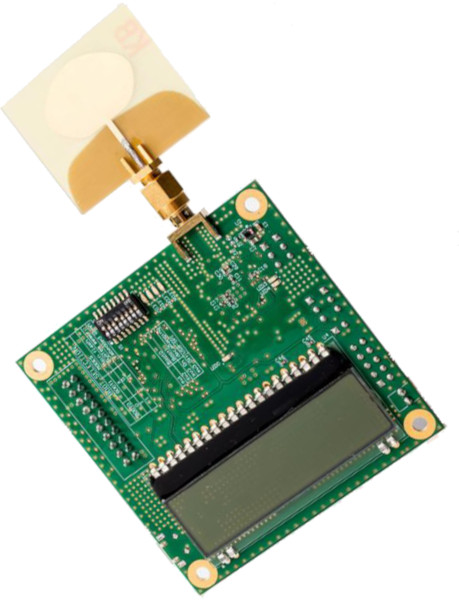}
		\label{img:evk1000}
	}
	\hfill
	\subfloat[MDEK1001 kit.]{%
		\includegraphics[height=2.25cm]{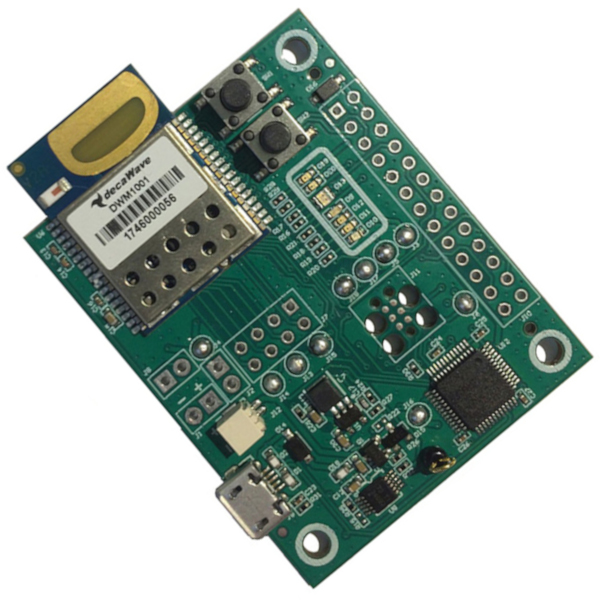}
		\label{img:dwm1001}
	}
	\hfill
	\subfloat[An antenna.]{%
		\includegraphics[height=3.0cm]{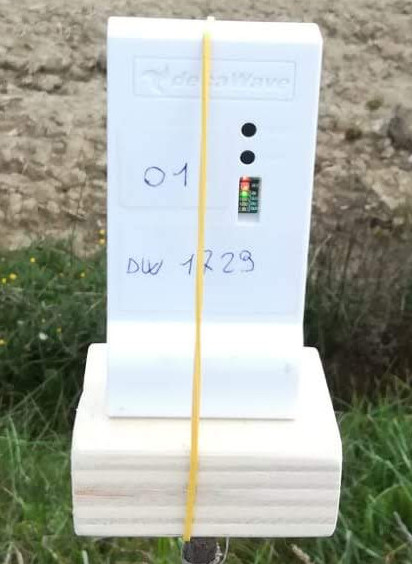}
		\label{img:exp-antenna}
	}
	\vspace{-0.1in}
	\hfill
	\subfloat[The drone.]{%
		\includegraphics[height=3.0cm]{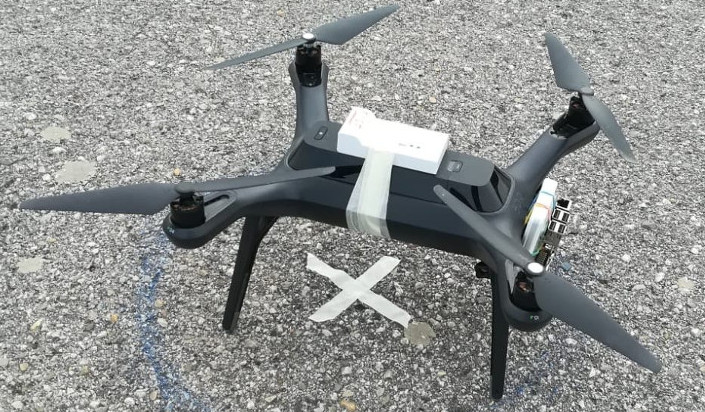}
		\label{img:exp-drone}
	}
	\vspace{-0.07in}
	\caption{The used \dw kits and the 3DR Solo drone.}
	\label{fig:kits}
	\vspace{-0.2in}
\end{figure}
}

\vspace{-0.05in}
\subsection{Experiments with Ground Error}\label{sec:1st-experiments}
In this section we analyze the ground error employing the \dw EVK1000 kit.
We start with pre-arranged antenna experiments in which two antennas
(one reproduces the \gd to localize and another one the \ma) are used.
The antenna that acts as \gd is fixed on the ground, while the other one (\ma),
fixed on a pole,
moves accordingly to the specific experiment emulating the rolling and altitude error.
\revision{We fixed the drone's position on the ground and that of the \gd measuring
the distance with a Bosch digital laser~\cite{www-laser}. 
The ground GPS drone's position is then converted as the origin $W'=(0,0,0)$
of the local Cartesian coordinate system used during the experiments.}
Then, in the subsequent experiments, we replace the pole  with a drone hovering at a certain altitude. 
The goal is to understand how the drone impacts the measurement error.
For each experiment, we record at least $30$ slant distances and we finally determine 
the final computed value at the $95\%$ confidence level.

\begin{figure*}[htbp]
	\centering
	\subfloat[Instrumental error.]{%
		\includegraphics[scale=0.75]{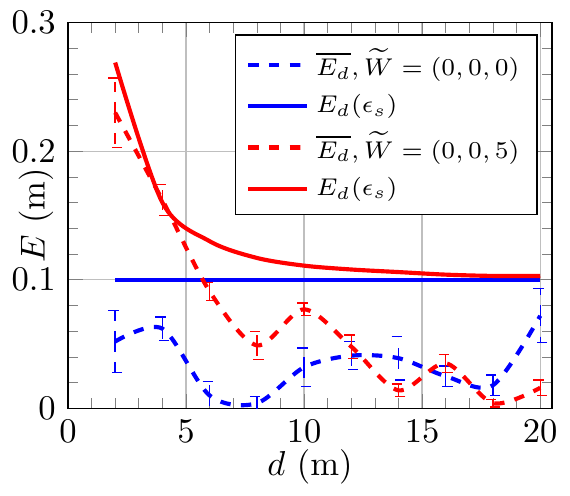}
		\label{fig:measurement_errors_instrumental}
	}
	\subfloat[Rolling error.]{%
		\includegraphics[scale=0.75]{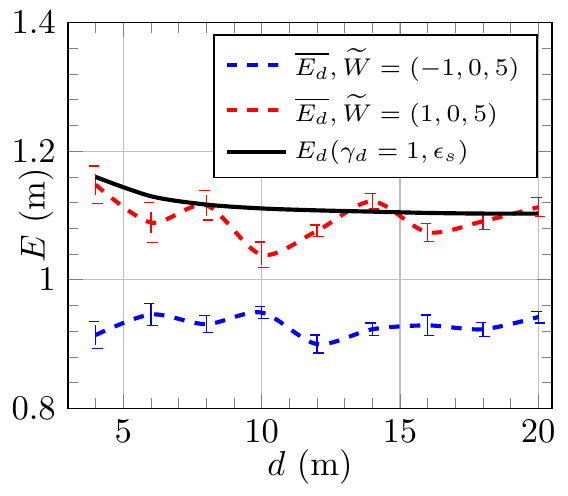}
		\label{fig:measurement_errors_rolling}
	}
	\subfloat[Altitude error.]{%
		\includegraphics[scale=0.75]{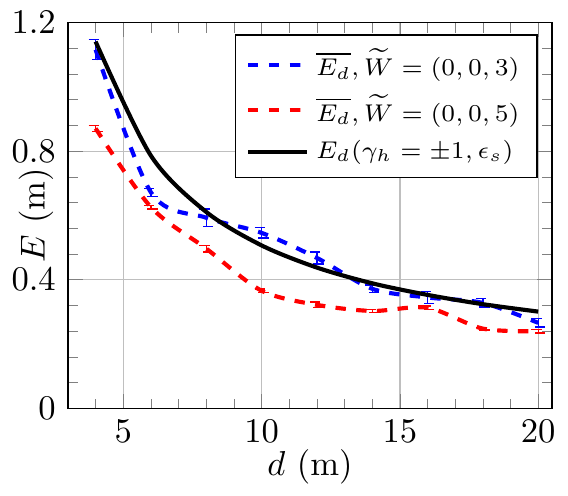}
		\label{fig:measurement_errors_altitude}
	}
	\subfloat[Combined error.]{%
		\includegraphics[scale=0.75]{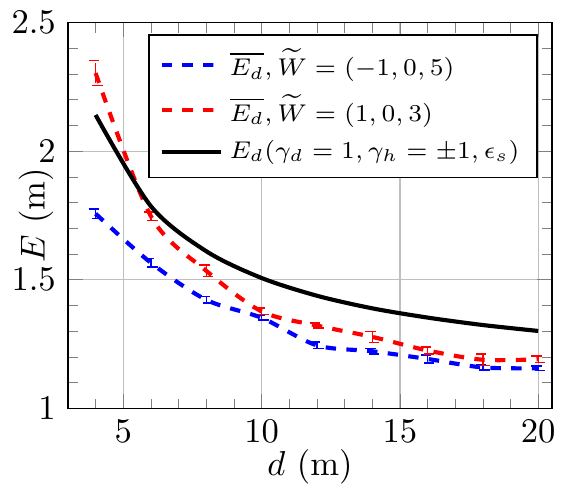}
		\label{fig:measurement_errors_combined}
	}
	\hfill
	\vspace{-0.1in}
	\subfloat[Experimental $\overline{E_d}$.]{%
		\includegraphics[scale=0.75]{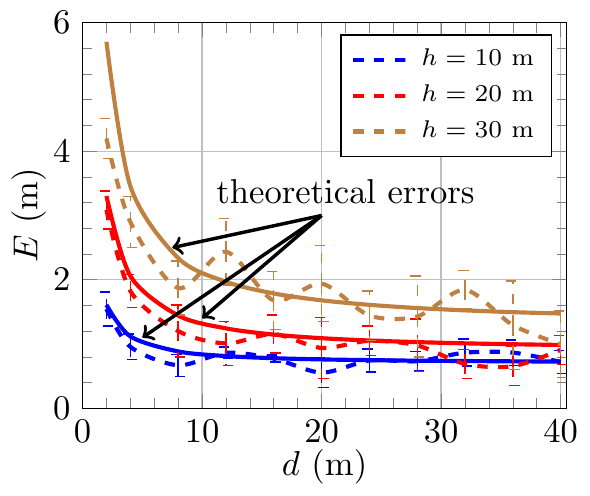}
		\label{fig:measurement_errors_drone}
	}
	\subfloat[Experimental $\overline{E^T_L}$: $d$ varies.]{%
		\includegraphics[scale=0.75]{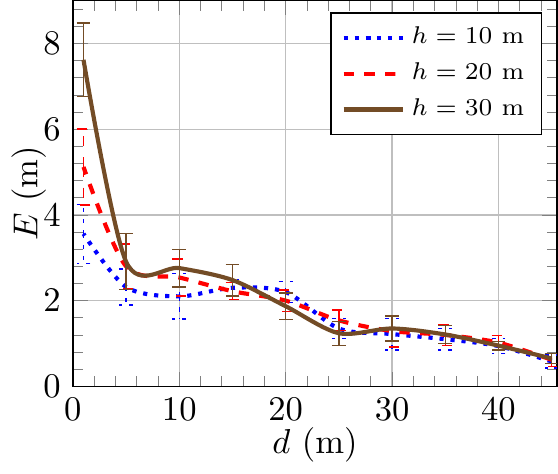}
		\label{fig:compare_measurements_same_angle}
	}
	\subfloat[Experimental $\overline{E^T_L}$: $\bmin$ varies.]{%
		\includegraphics[scale=0.75]{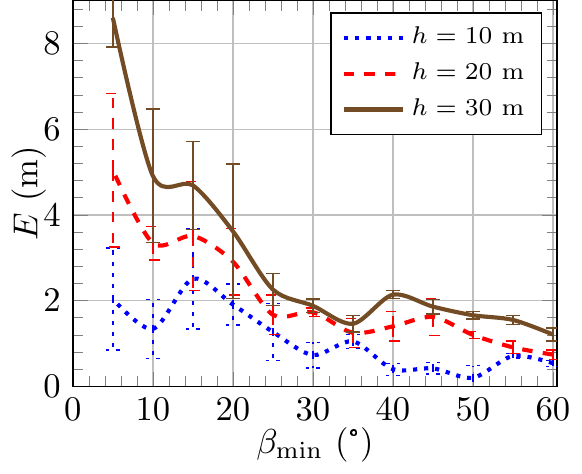}
		\label{fig:compare_measurements_same_distance}
	}
	\caption{The experimental error and the theoretical error in different cases.}
	\label{fig:compare_real_precision}
	\vspace{-0.1in}
\end{figure*}

In the first experiment, we measure the slant distance and
comparing its projection on the ground with the exact ground distance $d$.
Consequently, we compute the {\em experimental ground error} $\overline{E_d}$ 
and compare it with theoretical error $E_d$.
To verify Eq.~\eqref{eq:ground_distance_instrumental},
Eq.~\eqref{eq:ground_distance_gamma_d_roll_away}, and
Eq.~\eqref{eq:ground_distance_gamma_h_bound},
we have measured and reported in
Figure~\ref{fig:measurement_errors_instrumental},
Figure~\ref{fig:measurement_errors_rolling}, and
Figure~\ref{fig:measurement_errors_altitude}
the experimental error $\overline{E_d}$ when 
the instrumental error, rolling error, and altitude error, separately affect the \gd on the ground,
respectively.
We also report the theoretical ground error bound $E_d$.
It is interesting to see that, in each plot, 
the measurement error $E_d$ (solid line) 
almost always upper-bounds the experimental error $\overline{E_d}$ (dashed line).
We also measured and reported in Figure~\ref{fig:measurement_errors_combined}
the combined error where all the three components affect the error
along with the bound in Eq.~\eqref{eq:ground_distance_combined_bound_def}. 
The curves almost coincide.

In the second experiment, we repeat the previous setting employing this time a drone.
In Figure~\ref{fig:measurement_errors_drone} 
we report the experimental $\overline{E_d}$ for different altitudes.
Since the drone's position is affected by the wind, air density, humidity, the strength of
the propellers, and GPS error, 
we know that even the slant distance is affected at the same time.
Moreover, we know from Eq.~\eqref{eq:ground_distance_combined_bound_def} that 
the error $\overline{E_d}$ increases when $h$ increases and when $d$ tends to $0 \unit{m}$.
In Figure~\ref{fig:measurement_errors_drone}, we also plot the theoretical error $E_d$
in solid lines,
fixing $\epsilon_s=0.10 \unit{m}$ and
using $\gamma_d=\{0.6, 0.8, 1.2\} \unit{m}$ and $\gamma_h=\{0.1, 0.15, 0.2\} \unit{m}$ 
for each $h=\{10, 20, 30\} \unit{m}$, respectively,
that empirically fit the experimental curves.
Differently from the previous ones,
this is the first experiment that somehow simulates a real scenario.
It is interesting to note that,  we can model the curve of the combined error
even in a non-optimal scenario, just tuning in advance the parameters
on Eq.~\eqref{eq:ground_distance_combined_bound_def}, which
provides a good approximation of the error.

In conclusion, upon this first campaign of experiments,
we can confirm that the measurement error is small when either
the ground distance between the drone and the \gd is large
or the altitude of the drone is low.

\vspace{-0.1in}
\subsection{Experiments on the Trilateration Error}\label{sec:2nd-experiments}
In this section, we describe two more comparative experiments 
to better understand how the localization error can be affected
when the trilateration method is applied.
From Eq.~\eqref{eq:eps_l}, it is clear that the localization error $E_L$ can be bounded
if the three waypoints are sufficiently apart from the \gd.
In other words, the three points must respect good geometry and minimum distance constraints.
\revision{In both the experiments we use our 3DR Solo drone as a \ma and placed a single \gd in $P=(0, 0, 0)$.
Moreover, the drone's initial position $W'$ 
was initially set at the same \gd's position, i.e., in $W' = P$.}

In the first experiment depicted in Figure~\ref{fig:compare_measurements_same_angle},
we plot the {\em experimental localization trilateration error} $\overline{E^T_L}$ between the estimated and the actual position of the \gd.
Here, we fix the best possible minimum angle $\bmin=60^{\circ}$ and we decrease the value of
the ground distance $d$ to smaller values.
For each value of $d$, we perform trilateration using three points which satisfy the optimal geometry.
As expected and according to Eq.~\eqref{eq:eps_l},
$\overline{E_L^T}$ is high when $d$ is short,
even though the minimum angle is fixed at the best possible value $60^{\circ}$.

In the second experiment shown in Figure~\ref{fig:compare_measurements_same_distance}
we do the opposite by
keeping a large and good enough ground distance $d=40 \unit{m}$,
and decreasing the value of \bmin to narrow values.
Even here we perform trilateration and according to Eq.~\eqref{eq:eps_l}, 
the error decreases when $\bmin$ increases.

\vspace{-0.05in}
\subsection{Comparison of Localization Algorithms}\label{sec:3rd-experiments}
In this section, we describe the hardware and software architecture of the comparative testbed. The goal is to evaluate the performance of different localization algorithms in-field.

In this testbed, we cannot use the previous EVK1000 kit since it is formed by
only two antennas, hence it is definitively not sufficient for evaluating a real scenario
in which we have to localize multiple \gds at once.
Instead, we move towards the larger set of antennas
relying on the new MDEK1001 kit from \dw, since it comprises of a set of twelve antennas. 
In addition, the testbed consists also of a Raspberry Pi which is the main component that 
auto-pilots the drone via Wi-Fi and sends UWB commands via 
a single UWB antenna that is physically connected to it by the serial peripheral interface (SPI).

\vspace{-0.05in}
\subsubsection{Testbed setup}
\begin{figure}[htbp]
\vspace{-0.1in}
	\centering
	\hfill
	\subfloat[The \ma and \gd $P$.]{%
		\def\svgscale{0.8}
		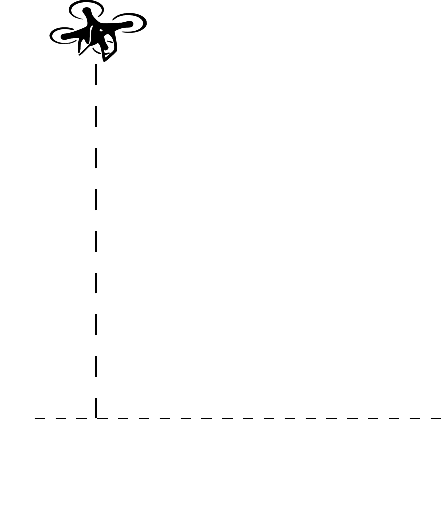
		\label{fig:ranging_precision_setup}
	}
	\hfill
	\subfloat[The deployment area.]{%
		\def\svgscale{0.5}
		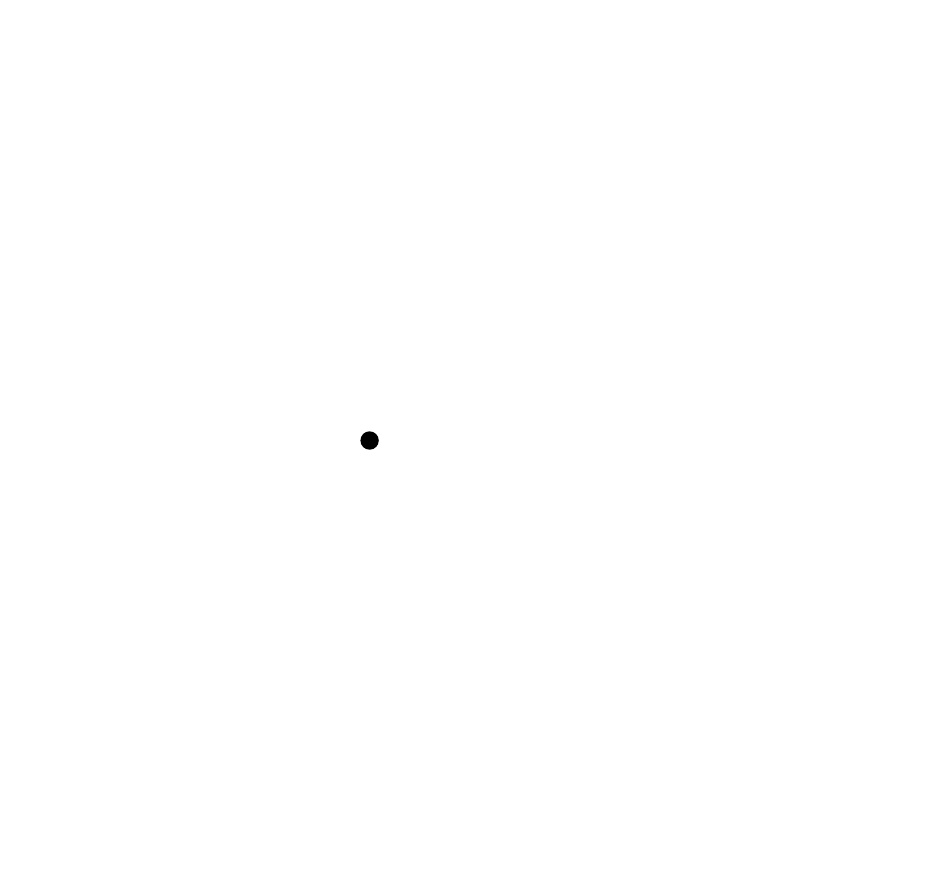
		\label{fig:deployment_area}
	}
	\caption{The experimental testbed on the field.}
	\label{fig:exps}
	\vspace{-0.1in}
\end{figure}

We set a rectangular deployment area of sizes $100 \times 100 \unit{m^2}$, and fix a Cartesian coordinate system with origin at 
the special position \home  $(0, 0, h_0 = 1 \unit{m})$.
Then, we deploy on the ground $n=10$ antennas placed at the top of a tripod of height $h_0$.
Each antenna  (Figure~\ref{img:exp-antenna}) identified by own ID is not aware of its relative
position with respect to the \home, even though we already know its position.
\revision{In fact, as illustrated in Figure~\ref{fig:deployment_area}, 
the deployed 10 antennas respect a predefined pattern, i.e., 
form a series of equilateral triangles (shown in green) with the same side of length $30 \unit{m}$
in which each vertex is a \gd.
Thus, we are able to measure, 
with reasonable accuracy before our experiments, 
the relative distance between the \gds, with the help of a digital laser.
Finally, the \home position is set between antennas ID 4 and ID 5 accurately measured with the same digital laser.}

By a drone's {\em mission} at a certain altitude $h$,
we actually refer to a drone (see Figure~\ref{img:exp-drone}) that flies at a fixed altitude $h_0+h$ (see Figure~\ref{fig:ranging_precision_setup}) following a certain static path \stp.
For each algorithm, the trajectory \stp
starts and finishes at \home and consists of vertical scans connected by horizontal scans (see Figure~\ref{fig:deployment_area}).
Once all the \gds are deployed, the drone starts its mission flying over the deployment area.
When both \gd and \ma are within the communication range of the other, 
the devices start a ToA based distance measurement protocol.
Then, the \gd stores the computed distance along with the current \ma's position.
In other words, the \gd memorizes the position of the waypoint and the associated distance on it.
At the end of the mission, each \gd estimates its position by invoking a localization algorithm.

All the compared algorithms require at least three points (see Table~\ref{tab:algorithms_evaluation})
to compute and estimate the position of a \gd.
However, each \gd has several stored distance measurements,
so it can potentially exploit all of them.
In order to better understand how either the altitude of the drone or the geometry of the waypoints impact
the quality of the localization \revision{accuracy}, 
as already investigated in Section~\ref{sec:2nd-experiments},
we fix two constraints during the selection of the three points:
\begin{inparaenum}[(i)]
	\item the ground distance $d$ between the \gd and the \ma, and
	\item the geometry angle $\beta$ to keep between the three waypoints.
\end{inparaenum}
Accordingly, we fix $d=\{20, 30, \ldots, 60\} \unit{m}$ and $\beta=\{0, 15, 30\}^{\circ}$,
where $\beta = 0^{\circ}$ means an unconstrained geometry.
Moreover, we vary the altitude $h=\{10, 20, 30\} \unit{m}$.
Clearly, it is not easy to find three points at an exact distance $d$.
Thus, we relax the constraint and we search for three points at distance $d \pm \tau$, 
where $\tau$ indicates a tolerance in our measurements (we fix $\tau=1\unit{m}$)
due to the fact that the drone sends its position at discrete intervals of time, i.e., 
the inter-waypoint distance \iw.
The \iw value is affected by the drone's speed.
In our experiments, we have seen that  $\iw=1 \unit{m}$ with a drone's speed of $10 \unit{m/s}$.

\subsubsection{Results}
We compare all algorithms varying the drone's altitude $h$, the minimum
distance $d$ among the \gds and the waypoints,
and the waypoint geometry. 
In \omni, by construction, we select always the furthest three waypoints that guarantee good geometry (see Figure~\ref{fig:angular_aperture}).
The \omni error is reported as a reference for the other algorithms. 

\begin{figure}[htbp]
\vspace{-0.15in}
	\subfloat[$h=10 \unit{m}, \beta = 0^{\circ}$.]{%
		\includegraphics[scale=0.75]{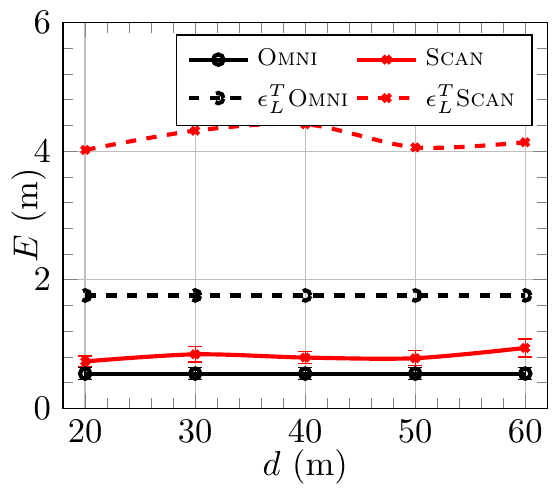}
		\label{fig:comparison_algorithms_trilat_h10_a0}
	}
	\subfloat[$h=10 \unit{m}, \beta = 30^{\circ}$.]{%
		\includegraphics[scale=0.75]{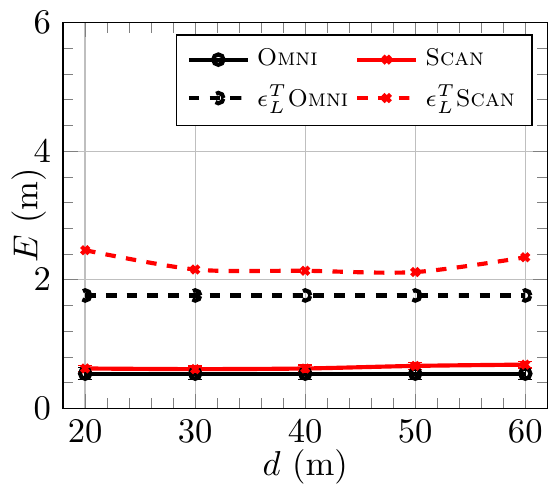}
		\label{fig:comparison_algorithms_trilat_h10_a30}
	}
	\hfill
	\vspace{-0.1in}
	\subfloat[$h=30 \unit{m}, \beta = 0^{\circ}$.]{%
		\includegraphics[scale=0.75]{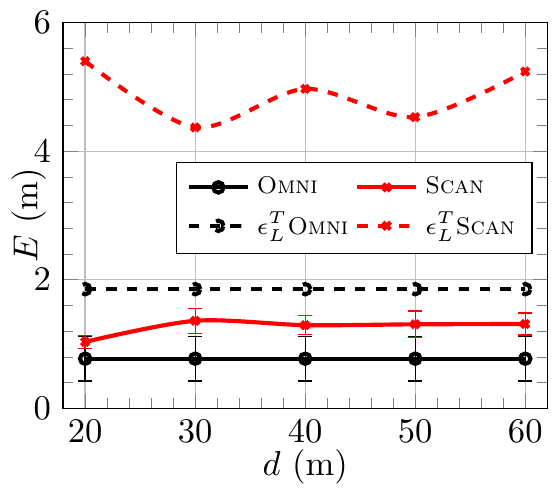}
		\label{fig:comparison_algorithms_trilat_h30_a0}
	}
	\subfloat[$h=30 \unit{m}, \beta = 30^{\circ}$.]{%
		\includegraphics[scale=0.75]{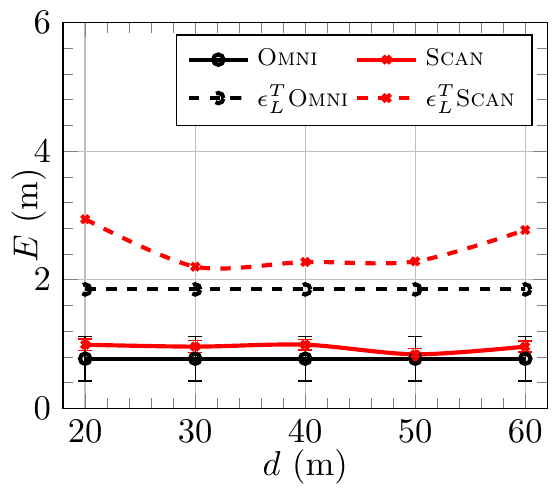}
		\label{fig:comparison_algorithm_trilats_h30_a30}
	}
	\caption{Errors for \scan and \omni.}
	\label{fig:comparison_algorithms_tri_err}
	\vspace{-0.05in}
\end{figure}

In Figure~\ref{fig:comparison_algorithms_tri_err}, we show the observed errors of
\scan and  \omni along with their
theoretical bounds $\epsilon^T_L$ given in Eq.~\eqref{eq:eps_l}
obtained by substituting  $\epsilon_s=0.10 \unit{m}$, and the values of $\gamma_d$ and $\gamma_h$
taken from Figure~\ref{fig:measurement_errors_drone}.
Obviously, \omni is better than \scan because the geometry of the waypoints is enforced.
The difference between the theoretical and the observed error is smaller for \omni than \scan
when $\beta=0^{\circ}$, and almost the same when $\beta=30^{\circ}$.

\begin{figure}[htbp]
\vspace{-0.15in}
	\subfloat[$h=30 \unit{m}, \beta = 0^{\circ}$.]{%
		\includegraphics[scale=0.75]{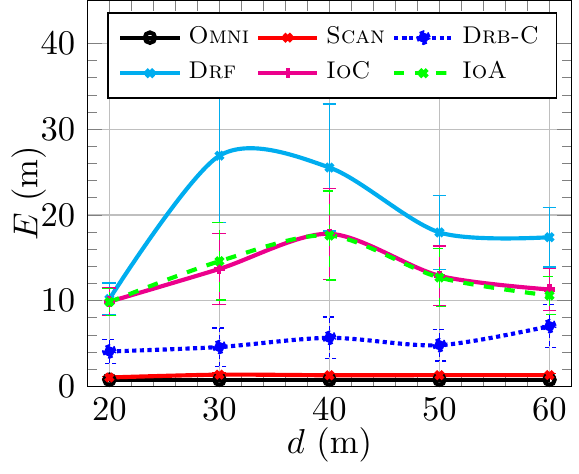}
		\label{fig:comparison_algorithms_h30_a0}
	}
	\subfloat[$h=30 \unit{m}, \beta = 30^{\circ}$.]{%
		\includegraphics[scale=0.75]{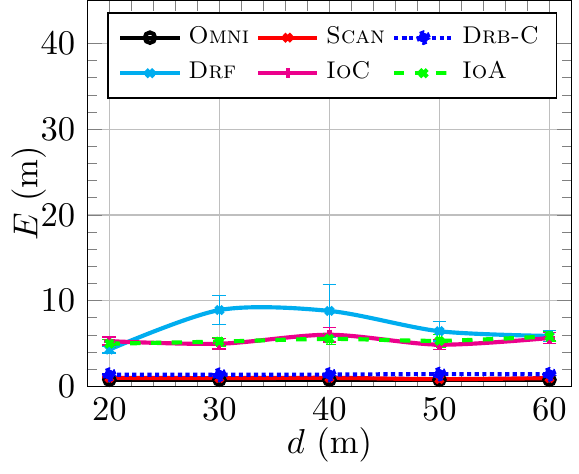}
		\label{fig:comparison_algorithms_h30_a30}
	}
	\hfill
	\vspace{-0.1in}
	\subfloat[$h=10 \unit{m}, \beta = 0^{\circ}$.]{%
		\includegraphics[scale=0.75]{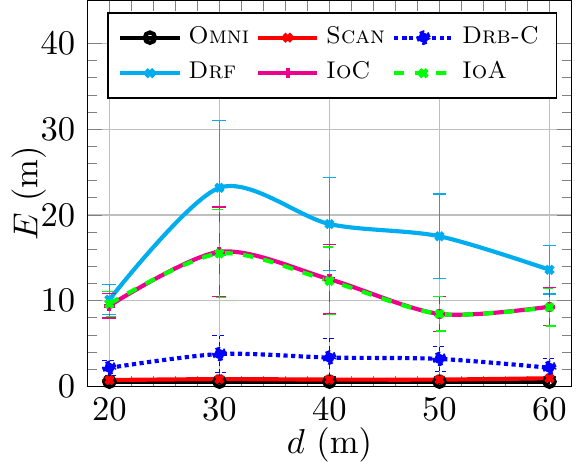}
		\label{fig:comparison_algorithms_h10_a0}
	}
		\subfloat[$h=10 \unit{m}, \beta = 30^{\circ}$.]{%
		\includegraphics[scale=0.75]{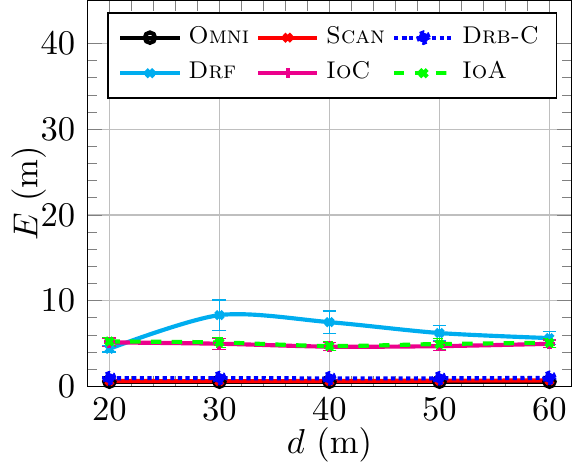}
		\label{fig:comparison_algorithms_h10_a30}
	}
	\caption{For fixed $h$ and $\beta$, the algorithms error when $d$ varies.}
	\label{fig:comparison_algorithms_loc_err}
	\vspace{-0.1in}
\end{figure}

\revision{Figure~\ref{fig:comparison_algorithms_loc_err} compares the  localization errors $E_L$ of the algorithms
when $d$ varies.
The localization error of \drf, \xiao, and \lee is greater than that of \scan and \omni.
As said, the trilateration based algorithms, \omni and \scan, get a good localization but
pose many constraints (angle, distance) in the selection of waypoints;
they also compute the estimated position by
performing a least-squares-error optimization technique (which is complex).
The \drbc is not as accurate as \omni or \scan because it omits the least-squares-error optimization technique, but it is quite good.
The chords based method in \drf is the least \revision{accurate}.
\xiao and \lee  improve over \drf because their localization technique
use the radius information.
The errors are large when $\beta=0^{\circ}$, while they significantly decrease for all the algorithms
when $\beta=30^{\circ}$. This shows that all the algorithms, and not only those based
on trilateration, benefit from a good waypoint geometry.
The experiments with $h=10 \unit{m}$ in Figures~\ref{fig:comparison_algorithms_h10_a0} and~\ref{fig:comparison_algorithms_h10_a30}  have a smaller elevation
angle and thus a smaller error than those with $h=30\unit{m}$
reported in Figures~\ref{fig:comparison_algorithms_h30_a0} and~\ref{fig:comparison_algorithms_h30_a30}.
When $h=30 \unit{m}$, all the ratios yield $h/d \ge 1/3$, and since our experiments are in sub-urban area
all the measurements are possibly in LoS;
whereas when $h=10 \unit{m}$,  the measurements at $d \ge 30\unit{m}$ are mixed LoS and NLoS.
Nonetheless, we cannot notice any special behavior, probably, 
thanks to the UWB multipath immunity.

\begin{figure}[htbp]
\vspace{-0.15in}
	\centering
	\subfloat[$h=10 \unit{m}, d = 30  \unit{m}$.]{%
		\includegraphics[scale=0.75]{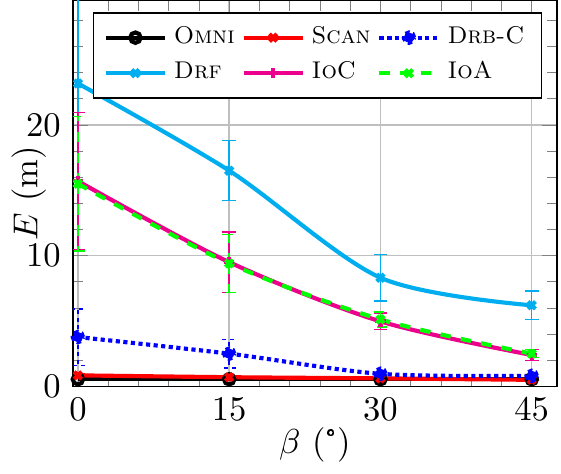}
		\label{fig:comparison_algorithms_h10_d30}
	}
	\subfloat[$h=10 \unit{m}, d = 50  \unit{m}$.]{%
		\includegraphics[scale=0.75]{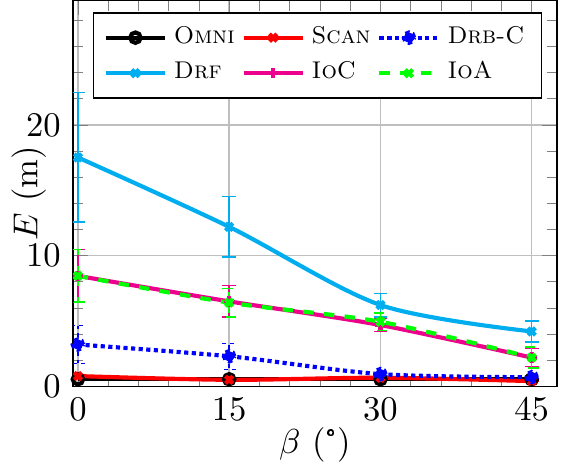}
		\label{fig:comparison_algorithms_h10_d50}
	}
	\caption{For fixed $h$ and $d$, the algorithms error when $\beta$ varies.}
	\label{fig:fixed_h_d}
\end{figure}

Figure~\ref{fig:fixed_h_d} compares the 
localization error $E_L$ when $h=10  \unit{m}$ and $d=30 \unit{m}$ or $d=50 \unit{m}$, varying
the geometry angle $\beta$.
From the observed errors, any localization that satisfies  $\beta \ge 30^{\circ}$ has a small
error, and cannot significantly improve decreasing the elevation angle (i.e., ratio $h/d$).
The decrease of the error when $h$ decreases from $50 \unit{m}$ to
$30 \unit{m}$ is large when $0^{\circ} \le \beta \le 30^{\circ}$.
Finally, note that in Figure~\ref{fig:comparison_algorithms_h10_d50},
it holds $h/d=0.2$, which is below the ratio that guarantees $100\%$
LoS in sub-urban area in Table~\ref{tab:lap}, but
we do not notice a meaningful worsening of the error.

Figure~\ref{fig:fixed_ratio} plots the
error for different pairs of $h$ and $d$ with the same ratio $h/d$.
Precisely, we compare two ratios $h/d$: $0.5$ and $1.0$.
Each ratio can be extracted from three different combinations altitude/distance.
For example, for $h/d=0.5$ we consider the combinations
$h=\{10, 20, 30\} \unit{m}$ and $d=\{20, 40, 60\} \unit{m}$.
The improvement in the accuracy is high when the elevation angle
decreases from $45^{\circ}$ to $26^{\circ}$.
\drf, the least accurate algorithm in all our experiments, is very sensitive
to the change of the elevation angle.

\begin{figure}
\vspace{-0.2in}
	\centering
	\subfloat[$h/d = 1.0$.]{%
		\includegraphics[scale=0.75]{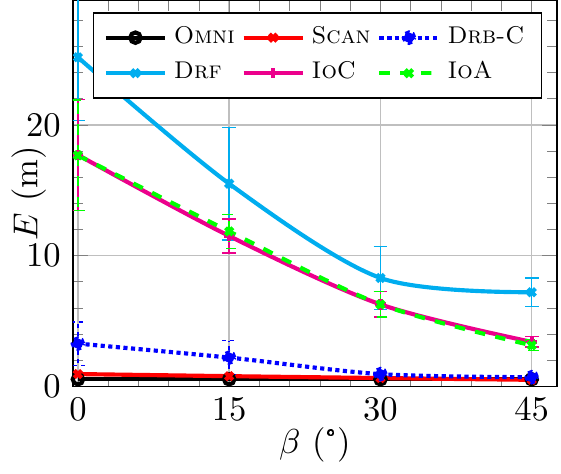}
		\label{fig:comparison_algorithms_hd_10}
	}
	\subfloat[$h/d = 0.5$.]{%
		\includegraphics[scale=0.75]{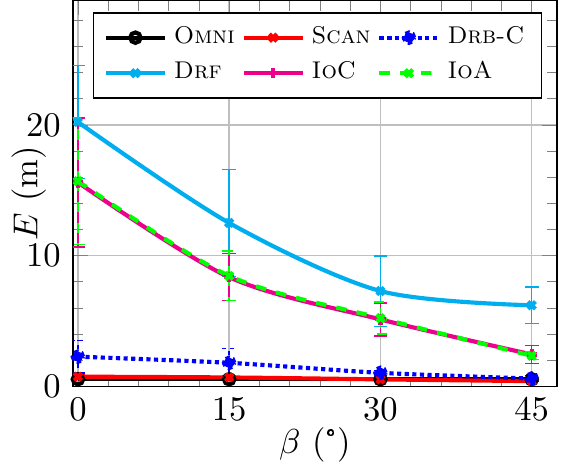}
		\label{fig:comparison_algorithms_hd_05}
	}
	\caption{For fixed $h/d$ ratio, the algorithms error when $\beta$ varies.}
	\label{fig:fixed_ratio}
\end{figure}
}

\begin{table}[ht]
\vspace{-0.1in}
	\renewcommand{\arraystretch}{1.15}
	\caption{Error between range-free and range-based algorithms, in meters (m).}
	\label{tab:comparison_algorithms_rf_rb}
	\centering
	\vspace{-0.1in}
	\begin{tabular}{ll|cccccc}
		\hline
		\multicolumn{2}{c}{ } & \multicolumn{2}{c}{\drf} & \multicolumn{2}{c}{\xiao} & \multicolumn{2}{c}{\lee} \\
		\multicolumn{2}{c}{ } & RF & RB & RF & RB & RF & RB \\
		\hline
		\multirow{3}{*}{$h$ (m)} & $10$ & $52.3$ & $16.6$ & $48.8$ & $10.8$ & $47.2$ & $10.8$ \\
		& $20$ & $55.1$ & $19.8$ & $49.1$ & $11.2$ & $48.8$ & $11.0$ \\
		& $30$ & $57.7$ & $20.4$ & $48.2$ & $13.8$ & $51.2$ & $13.6$ \\
		\hline
	\end{tabular}
	\vspace{-0.05in}
\end{table}

\revision{In conclusion,} in Table~\ref{tab:comparison_algorithms_rf_rb}
compares the localization error $E_L$ among range-free (RF) and range-based (RB) versions
of the three original range-free algorithms \drf, \xiao, and \lee, for different altitudes $h$.
In particular, we report the localization error obtained from our previous testbed~\cite{bettisorbelli2020rangefree}
on which the three algorithms were implemented as pure range-free techniques
based on the HnH technique (RF columns),
along with the average results shown in Figure~\ref{fig:comparison_algorithms_loc_err} (RB columns).
As reported in~\cite{bettisorbelli2020rangefree}, on average, the experimental error of those algorithms is very large (almost $60 \unit{m}$) and variable.
These experiments show that the error of the original range-free versions
is 3-4 times larger than the corresponding extended range-based version 
that exploit distance measurements.
Moreover, in~\cite{bettisorbelli2020rangefree}, about one-third 
of the \gds were left unlocalized by \xiao and \lee,
while \drf localized all of them.
These results and the fact that our antennas are able to take distance measurements via ToA
fully justify our transformation of \drf, \xiao, and \lee in range-based algorithms, 
implying that measurements help.

\vspace{-0.1in}
\section{Conclusions}\label{sec:concl}
In this paper, we analytically study and experimentally evaluate thorough real experiments on the field the
errors that can affect the localization of \gds using a drone as \ma.
We decompose the error in measurement error, ground error, and localization error, 
and provide analytical expressions of these errors. 
We also link the ground error with the theory of the A2G communication link via the elevation angle.
Our experiments confirm that our analytical analysis is accurate. 
Furthermore, results also show that extending range-free algorithms with range-based measurements, 
significantly increases the localization \revision{accuracy}. 

\revision{In the future, we plan to extend the analysis to NLoS scenarios.
We will investigate \dw antenna capabilities, 
and also modulating and tuning the transmitting power in different scenarios.
We finally plan to extend our work to propose a more realistic antenna radiation pattern.
}

\vspace{-5pt}
\paragraph{Acknowledgments}
The authors are grateful to the editor and reviewers for valuable comments that helped us improve the quality of the manuscript.
This work was partially supported by Project {\em NALP-SAPR} granted by FSE,  
Project {\em NALP-SAPR2} granted by University of Perugia, by NATO grant G4936, 
Intelligent Systems Center (ISC) at Missouri S\&T,
and by NSF grants CNS-1545050, CNS-1725755, CNS-1818942, and SCC-1952045.

\vspace{-0.05in}
\bibliographystyle{IEEEtran}
\bibliography{IEEEabrv,main}

\vspace{-0.3in}
\begin{IEEEbiography}
[{\includegraphics[width=1in,height=1.25in,clip,keepaspectratio]{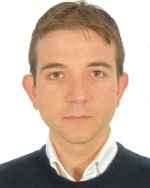}}] {Francesco Betti Sorbelli}
received the Bachelor and Master degrees {\em cum laude} in Computer Science from the University of Perugia, Italy, in 2007 and 2010, respectively, 
and his Ph.D. in Computer Science from the University of Florence, Italy, in 2018.
After his Ph.D., he was a Postdoc Researcher at University of Perugia.
Currently, he is a Postdoc at the
Missouri University of Science and Technology University, USA.
His research interests include wireless sensor networks,
algorithms on drones and robots.
\end{IEEEbiography}

\vspace{-0.3in}
\begin{IEEEbiography} [{\includegraphics[width=1in,height=1.25in,clip,keepaspectratio]{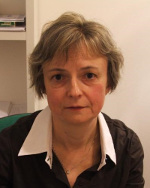}}] {Cristina M. Pinotti}
received the Master degree {\em cum laude} in Computer Science from the University of Pisa, Italy, in 1986.
In 1987-1999, she was Researcher with the National Council of Research in Pisa.
In 2000-2003, she was Associate Professor at the University  of Trento.
Since 2004, she is a Full Professor at the University of Perugia.
Her current research interests include the design and analysis of algorithms for wireless sensor networks and communication networks.
She has published  more than 100 refereed papers on international journals and conferences, workshops, and book chapters.
\end{IEEEbiography}

\vspace{-0.3in}
\begin{IEEEbiography} [{\includegraphics[width=1in,height=1.25in,clip,keepaspectratio]{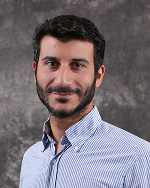}}] {Simone Silvestri} graduated with honors and received his PhD in
computer science at Sapienza University of Rome, Italy. He is an
Assistant Professor in the Department of Computer Science of the
University of Kentucky. His research interests include
cyber-physical-human systems, Internet of Things, hybrid sensor
networks, and network management. He servers as Associate Editor of
Elsevier's Pervasive and Mobile Computing journal, and has served on the
Technical Program Committees of numerous international conferences. His
research is funded by several agencies such as NIFA, NSF, and NATO. He
received the NSF CAREER award in 2020.
\end{IEEEbiography}

\vspace{-0.2in}
\begin{IEEEbiography} [{\includegraphics[width=1in,height=1.25in,clip,keepaspectratio]{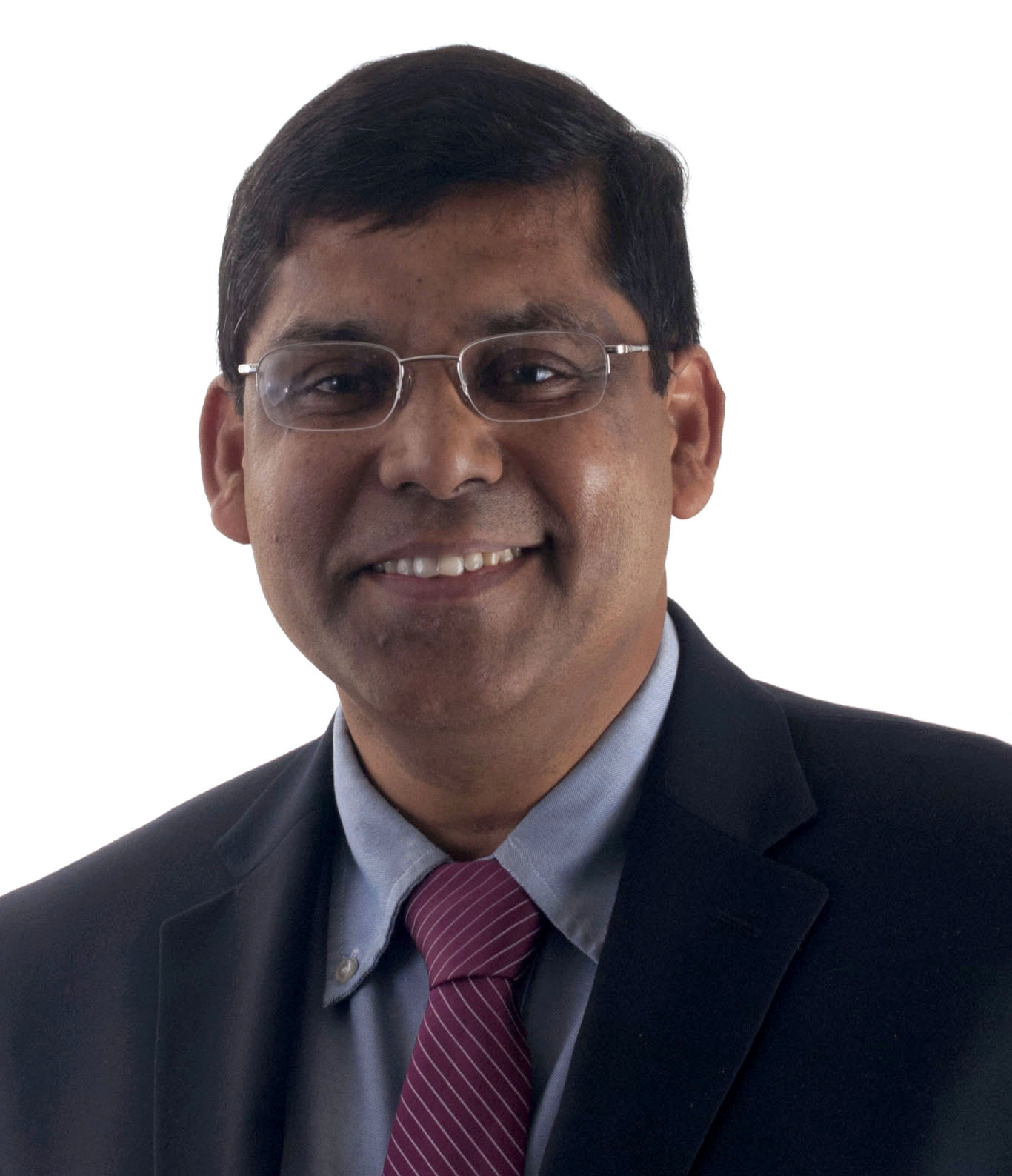}}] {Sajal K. Das} is a professor of computer science and Daniel St. Clair Endowed Chair at Missouri University of Science and Technology. 
His research interests include wireless sensor networks, mobile and pervasive computing, cyber-physical systems and IoT, smart environments, cloud computing, cyber security, and social networks. He serves as the founding Editor-in-Chief of Elsevier's Pervasive and Mobile Computing journal, and as Associate Editor of several journals including the IEEE Transactions of Mobile Computing, IEEE Transactions on Dependable and Secure Computing, and ACM Transactions on Sensor Networks. He is an IEEE Fellow.
\end{IEEEbiography}

\end{document}